\documentclass[structabstract]{aa}
\usepackage{graphicx,natbib,amssymb}

\newcommand{\plotwd}{9.cm}
\newcommand{\plotwdtwo}{18.cm}

\newcommand{\cs}{\langle\sigma_A\upsilon\rangle}

\usepackage{hyperref}
\usepackage{breakurl}

\begin{document}
\title{WMAP7 and future CMB constraints on annihilating dark matter: implications for GeV-scale WIMPs}
\titlerunning{CMB constraints on annihilating DM}
\author{Gert H\"utsi \inst{1,2}, Jens Chluba \inst{3}, Andi Hektor \inst{4}, Martti Raidal \inst{4}}
\institute{Tartu Observatory, T\~oravere 61602, Estonia \\ \email{gert@aai.ee} \and Max-Planck-Institut f\"ur Astrophysik, Karl-Schwarzschild-Str. 1, D-85741 Garching, Germany \and Canadian Institute for Theoretical Astrophysics, 60 St. George Street, Toronto, ON M5S 3H8, Canada \\ \email{jchluba@cita.utoronto.ca} \and National Institute of Chemical Physics and Biophysics, Tallinn 10143, Estonia\\
\email{andi.hektor@cern.ch, martti.raidal@cern.ch}}
\date{Received / Accepted}

\abstract
{{We calculate constraints from current and future cosmic microwave background (CMB) measurements on  annihilating dark matter (DM) with masses below the electroweak scale: $m_{{\rm DM}}=5-100$ GeV. In particular, we assume the S-wave annihilation mode to be dominant, and focus our attention on the lower end of this mass range, as DM particles with masses $m_{{\rm DM}}\sim 10\,$GeV have recently been claimed to be consistent with the CoGeNT and DAMA/LIBRA  results, while also providing viable DM candidates to explain the measurements of Fermi and WMAP haze.}}
{We study the model (in)dependence of the CMB power spectra on particle physics DM models, large-scale structure formation and cosmological uncertainties. We attempt to find a simple and practical recipe for estimating current and future CMB bounds on a broad class of DM annihilation models.}
{We use a model-independent description for DM annihilation into a wide set of Standard Model particles simulated by PYTHIA Monte Carlo. Our Markov chain Monte Carlo calculations used for finding model constraints involve realistic CMB likelihoods and assume a standard 6-parameter $\Lambda$CDM background cosmological model, which is extended by two additional DM annihilation parameters: $m_{{\rm DM}}$ and $\cs/m_{{\rm DM}}$.}
{We show that in the studied DM mass range the CMB signal of DM annihilations is independent of the details of large-scale structure formation, distribution, and profile of DM halos  and other cosmological uncertainties. All particle physics models of DM annihilation can be described with only one parameter, the fraction of energy carried away by neutrinos in DM annihilation. As the main result we provide a simple and rather generic fitting formula for calculating CMB constraints on the annihilation cross section of light WIMPs. We show that thermal relic DM in the CoGeNT, DAMA/LIBRA favored mass range is in a serious conflict with present CMB data for the annihilation channels with few neutrinos, and will definitely be tested by the Planck mission for all possible DM annihilation channels. Also, our findings strongly disfavor the claim that thermal relic DM annihilations with $m_{{\rm DM}}\sim 10$ GeV and $\cs\sim 9\times10^{-25}$ ${\rm cm}^3{\rm s}^{-1}$ could be a cause of Fermi and WMAP haze.}
{}
\keywords{Cosmology: theory -- dark matter -- diffuse radiation -- cosmic microwave background -- Elementary particles}
\maketitle

\section{Introduction}
According to our current understanding, the energy-matter content of the Universe is dominated by dark components: dark energy (DE) ($\sim 73\%$) and dark matter (DM) ($\sim 22\%$), with ordinary baryonic matter accounting for only $\sim 5\%$ of the total ($\sim$ critical) density \citep[e.g., see][]{2011ApJS..192...18K}. In contrast to the poorly understood DE -- the substance causing the Universe to expand in an accelerated fashion -- we have physically well-motivated models for the DM. Among those, the most promising scenario states that the DM of the Universe consists of thermal relic density of stable weakly interacting massive particles (WIMPs). It is quite miraculous that having particles with masses and annihilation cross sections set by the electroweak scale automatically provide the right DM density after freeze-out \citep{1996PhR...267..195J,2005PhR...405..279B}. 

The WIMP hypothesis, along with its potentially observable phenomenology, has initiated strong effort in the particle- and astrophysics communities to try to find other than purely gravitational manifestations of DM. So far, we have good knowledge of DM only through its gravitational effects, starting from the scale of galaxies and galaxy clusters, up to the cosmologically largest observable scales \citep{1996PhR...267..195J,2005PhR...405..279B,2009arXiv0901.0632E}. However, as there is already an impressive list of ongoing and upcoming direct DM detection experiments along with various indirect means of detection \citep[see][for overview]{2010ARA&A..48..495F}, the hopes are very high that in the nearest future the mystery of DM might at last be solved. Indeed, the first signals from DM particles could potentially have already been detected: the (expected) annual modulation signal from DAMA/LIBRA \citep{2010EPJC...67...39B}, signals from the CDMS \citep{2011PhRvL.106m1302A} and CoGeNT \citep{2011PhRvL.106m1301A} nuclear recoil experiments, anomalies of the cosmic ray positrons as revealed by PAMELA satellite \citep{2009Natur.458..607A}, or positrons+electrons as obtained by the Fermi satellite \citep{2009PhRvL.102r1101A} and HESS atmospheric Cherenkov telescope \citep{2009A&A...508..561A}. While the cosmic ray positron anomaly can possibly be explained by TeV-scale DM \citep{2008PhRvD..78j3520B,2009PhLB..672..141B,2009NuPhB.813....1C,2009PhRvD..79a5014A,2009PhRvD..79h3528F}, the signal from CoGeNT calls for light WIMPs within the mass range $7-12$~GeV \citep{2011PhRvL.106m1301A}. Consistent analyses of combined data from CoGeNT and DAMA/LIBRA determine the light DM mass to be  $6-8$~GeV \citep{2010PhRvD..82l3509H}. Although this mass range is probed by the CDMS \citep{2011PhRvL.106m1302A}, XENON10 \citep{2009PhRvD..80k5005A}, and XENON100 \citep{2010PhRvL.105m1302A,2011arXiv1103.0303X} experiments, interpretation of those results \citep{2010JCAP...02..014K,2010arXiv1011.5432S} requires an ability to reliably reconstruct nuclear recoils at very low energy \citep{2011PhRvD..83e5002S}, as well as precise knowledge of DM distribution and velocity in the local halo.  Therefore the CoGeNT  and DAMA/LIBRA hints of light DM cannot be ruled out unambiguously.

In addition, there is an independent positive claim of the existence of {\cal O}(10)~GeV mass DM. A recent study by \citet{2011arXiv1102.5095D} also suggests that annihilating DM with similarly low masses ($m_{{\rm DM}}=1-20$ GeV) may give a good match to the observed Fermi and WMAP haze \citep{2010ApJ...717..825D,2008ApJ...680.1222D}. However, all those claims depend strongly on the knowledge of the profile of the DM halo of our Galaxy and precise knowledge of local DM density and halo substructure.

Thus the several interesting claims of the existence of {\cal O}(10)~GeV mass DM  call for model-independent tests of the light DM scenario.   Since lower DM particle masses imply higher number densities ($n_{\rm DM}\propto \Omega_{\rm DM}h^2/m_{{\rm DM}}$), and as the energy input from annihilations scales as $\propto n_{{\rm DM}}^2$, one might expect strong constraints on annihilation cross section, which might possibly reach below the standard thermal production value of $\cs\simeq 3\times10^{-26}$ cm$^3$s$^{-1}$ \citep{1996PhR...267..195J}. The constraints from gamma-ray measurements \citep{2009PhRvD..79d3507B,2009JCAP...03..009B,2009PhRvD..79h1303B,2009NuPhB.821..399C,2010NuPhB.831..178M,2010NuPhB.840..284C,2010JCAP...03..014P,2010JCAP...07..008H,2010PhRvD..82l3511B,2011JCAP...01..011A,2010PhRvD..82l3519V,2011PhRvD..83l3513Z}  along with CMB bounds \citep{2009PhRvD..80b3505G,2009PhRvD..80d3526S,2009JCAP...10..009C,2009A&A...505..999H,2010PThPh.123..853K}, indicate that this might indeed be the case. 

In this paper we investigate how well DM annihilation cross sections for WIMPs with masses $m_{{\rm DM}}=5-100$ GeV can be constrained with CMB measurements, in particular focusing on the lower end of this mass range. The main advantage of CMB over other indirect probes, like gamma ray and cosmic ray measurements, is that it is practically insensitive to the complications caused by the nonlinear evolution of the cosmic density field. As our analysis shows, for any realistic structure formation scenario the CMB bounds on annihilating DM arise solely around the redshifts of $z\sim 1000$, while the contribution from lower redshift cosmic structures is completely negligible. 

CMB constraints on annihilating DM have been obtained in several earlier studies: e.g. \citet{2005PhRvD..72b3508P,2006MNRAS.369.1719M,2006PhRvD..74j3519Z,2009PhRvD..80b3505G,2009PhRvD..80d3526S,2009JCAP...10..009C,2009A&A...505..999H,2010PThPh.123..853K}. Even though most of these analyses have assumed a simple `on the spot' approximation for the energy deposition\footnote{I.e., energy input from DM annihilations is assumed to get instantaneously absorbed by the cosmic medium, with some efficiency factor $f$.}, more recent studies \citep{2009PhRvD..80d3526S,2009A&A...505..999H} followed the energy transport problem including various energy-loss mechanisms in a more realistic way. Compared to the analysis of \citet{2009PhRvD..80d3526S}, which partially relies on the previously derived `on the spot' results of \citet{2009PhRvD..80b3505G}, in this paper we perform a more elaborate treatment for the energy deposition joined to the analysis of the CMB data via Markov chain Monte Carlo calculations that incorporate the most recent WMAP likelihood code. Also, we make an attempt to unify the results from various annihilation channels and provide a simple and rather generic fitting formula for calculating CMB constraints on annihilation cross section $\cs$ for a broad range of annihilating DM models. 

Our paper is organized as follows. In Section 2 we give a brief description of the energy input from DM annihilation and provide a simple treatment for its propagation. The effect on CMB temperature and polarization fluctuations is investigated in Section 3. Section 4 presents our main results about current and future CMB constraints. Our summary is given in Section 5.

\section{Input signals and their propagation}\label{sec2}

As in \citet{2009NuPhB.813....1C,2011JCAP...03..051C}, we treat our input signals from DM annihilation in an as model independent a way as possible. In \citet{2011JCAP...03..051C} the two-particle annihilation channels to all Standard Model (SM) particles were considered: leptons, quarks, photons, gluons, weak-interaction gauge bosons, Higgs boson, and neutrinos. In addition, annihilations to four leptons via an intermediate new boson $V$ were considered. Such a treatment can be considered as model independent since realistic models can always be decomposed into these basic channels where the particular branching ratios between the channels are given by the underlying theoretical particle physics model.
\begin{figure}
\centering
\includegraphics[width=\plotwd]{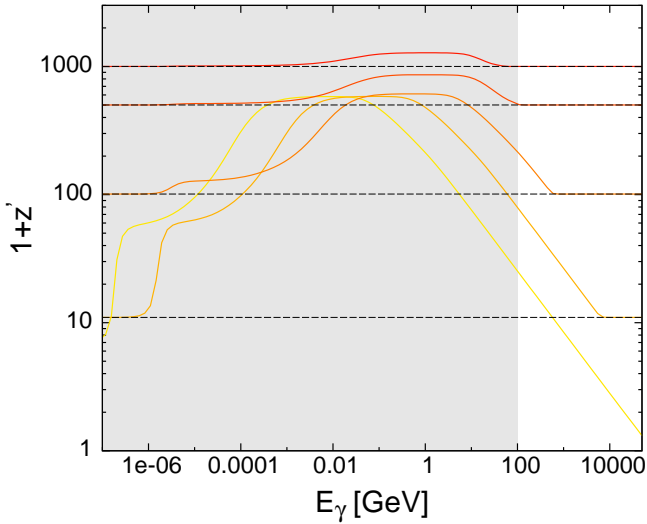}
\caption{The redshift $z^{'}$ where the optical depth for photons reaches unity (i.e., $\tau(z,z^{'})=1$) for several `observer's redshifts': $z=0,10,500,1000$. Here the energy plotted is the photon energy at redshift $z$. The light gray region corresponds to the DM mass interval considered in this paper.}
\label{fig1}
\end{figure}

Since in our work we focus on DM particle masses below $100$ GeV, out of all of the above channels the following remain: DM DM $\rightarrow$ SM $\overline{{\rm SM}}$, where SM$=\{e,\mu,\tau,q,c,b,\gamma,g$\}; plus 4-lepton channels via $V$. Here $q$ denotes the light quarks $u$, $d$, and $s$. Because the masses of interest in this work are mostly below the masses of the electroweak gauge bosons, $W^{\pm}$ and $Z$, those channels are left out. For the same reason we also do not need to distinguish between left- and right-handed particles. Also, the neutrino channels in this case provide only trivial output; i.e. $100\%$ of the energy is carried away by neutrinos, which escape freely at the redshifts of interest, so are not treated any further. Even though the channels $\gamma$ and $g$ are included in our model-independent approach, these are strongly suppressed for realistic models since DM should not carry color or interact electromagnetically.

For all channels, the spectra of the emerging stable particles, $e$, $p$, $\gamma$, $\nu$, after treatment of several decays, parton showers, and hadronization were calculated using PYTHIA Monte Carlo\footnote{\url{http://home.thep.lu.se/~torbjorn/Pythia.html}} \citep{2008CoPhC.178..852S}. All input spectra are downloadable from \url{http://www.marcocirelli.net/PPPC4DMID.html}. For more details, and in particular for a discussion on the level of possible uncertainties, we refer the reader to \citet{2011JCAP...03..051C}.

Among the stable output particles, neutrinos propagate freely at the redshifts of interest, while $e^{\pm}$ immediately interact with the ubiquitous CMB photons and upscatter those to the gamma-ray energy range via the inverse Compton (IC) mechanism. The total output energy in hadrons ($p$ and $d$) is typically quite negligible. Only in gluon and quark channels does it reach up to $\sim 15\%$, and that only for the highest DM particle masses considered in this paper. We therefore did not model this component in detail. However, in Sect.~\ref{sec:protons} we give some arguments that at the redshifts of interest, i.e. $z\sim 1000$, probably the majority of energy released in $p$ and $d$ channels is directly converted into heat by the cosmic medium. Thus, we only need a detailed treatment for the photons: (A) the energetic ones originating directly from the annihilation event (prompt photons), (B) the softer IC photons created by $e^{\pm}$ upscattering CMB photons. 

The processes and the corresponding cross sections relevant to the propagation of photons through the cosmic medium were taken from \citet{1989ApJ...344..551Z}. Starting from the lowest of energies these include (i) photoionization, (ii) Compton losses (on both bound and free electrons), (iii) pair production on matter, (iv) photon-photon scattering, and (v) pair production on ambient photon fields. 

In Fig.~\ref{fig1} we show the redshifts $z^{'}$ where the optical depth $\tau(z,z^{'})=1$ for various redshifts of the observer: $z=0,10,500,1000$. As can be seen, for intermediate photon energies (depending on the redshift) there is a well-known X-ray/gamma-ray energy window where the photons can propagate freely over cosmologically large distances \citep[e.g., see][]{2004PhRvD..70d3502C}. For the cosmological radiation transfer it is crucial that this `transparency window' is properly modeled. Once the photon gets outside of this window (we take it to happen after the first interaction) we assume that the following cascade will be locally absorbed in a very short time. Moreover, the fractions $(1-f_{{\rm ion}})/3$ and $(1+2f_{{\rm ion}})/3$ of the total absorbed energy, are assumed to be going for ionization and heating, respectively. Here $f_{{\rm ion}}$ is the fraction of ionized hydrogen atoms, and a similar expression for helium can be used. Excitations of atoms are neglected. This approximation, motivated by the work of \citet{1985ApJ...298..268S}, has been widely used in several subsequent papers, e.g. \citet{2004PhRvD..70d3502C,2005PhRvD..72b3508P,2006MNRAS.369.1719M,2006PhRvD..74j3519Z,2009PhRvD..80d3529N}. However, it is clear that these simple expressions only provide a rough estimate for the correct energy deposition efficiencies. As mentioned by \citet{2010MNRAS.402.1195C}, the precise redshift dependence of the efficiency factors and the fraction of energy that goes into excitations of hydrogen and helium atoms need more careful consideration of the radiative transfer processes, including secondary low-energy photons and their feedback. These extra photons have the potential of further delaying recombination, hence affect the last scattering surface and CMB anisotropies \citep{2000ApJ...539L...1P}. We leave a more detailed investigation of these ambiguities to a future paper; however, later on in Section~\ref{sec4} we briefly comment on how much the omission of excitations using the rough prescription of \citet{2004PhRvD..70d3502C} changes our final results.

In Fig.~\ref{fig1} we also see that at high redshifts, as the density of the environment becomes much higher, the X-ray/gamma-ray transparency window starts to close. Thus at sufficiently high $z$, we would expect all the produced annihilation energy (excluding the energy stored in neutrinos, since these can freely leak out) to be absorbed locally. In the following we call the ratio of the locally produced to the locally absorbed energy the $f$-parameter. At high redshifts (but well after the neutrino decoupling) we expect the $f$-parameter to asymptote to the value given by $(1-f_{\nu})$, where $f_{\nu}$ is the fraction of energy carried away by neutrinos. An example for the $f$-parameters in the case of $\mu$ annihilation channel are shown in the upper lefthand panel of Fig.~\ref{fig2}. Since $\sim 60\%$ of the energy is carried away by neutrinos in the $\mu$-channel the expected asymptotic high-redshift $f$-parameter should be $\sim 0.4$, which is indeed the case. With this in mind, we see that robust model-independent results from the CMB analyses can be obtained for the DM masses below $\sim 100$~GeV. This is the reason we concentrate on light WIMPS in this work. For heavier WIMPs, the computation depends on more complicated details for energy absorption. 

\section{Effect on CMB}

\begin{figure*}[ht!]
\centering
\includegraphics[width=18.5cm]{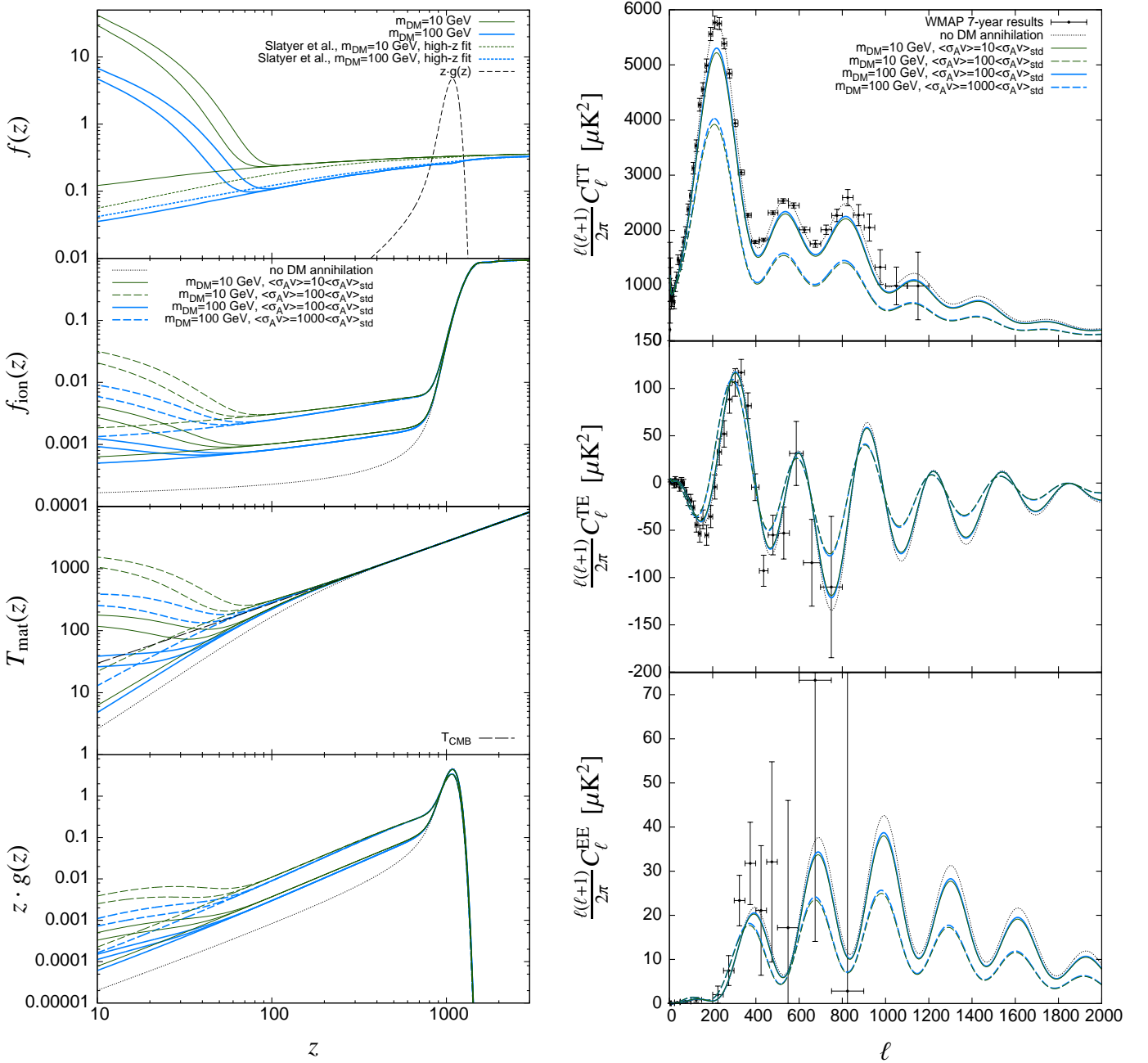}
\caption{{\bf Lefthand column} from top to bottom: {\bf (i)} $f$-parameters for the $\mu$-channel assuming $m_{{\rm DM}}=10$ GeV (dark solid lines) and $m_{{\rm DM}}=100$ GeV (light solid lines). In both cases, the lowest curve out of the triple of lines corresponds to only the smooth background, the middle one includes halos with a lower mass cutoff of $10^{-6}M_{\odot}$, while the top one has $10^{-9}M_{\odot}$. The dotted lines represent high-redshift fits (valid for $z\gtrsim 170$) as given by \citet{2009PhRvD..80d3526S}. The dashed line shows the CMB visibility function. Here $zg(z){\rm d}\ln z$ gives the probability that the CMB photon last scattered in the logarithmic redshift interval ${\rm d}\ln z$ centered on redshift $z$. {\bf (ii)} Fraction of free electrons as a function of redshift for $m_{{\rm DM}}=10$ GeV and $100$ GeV, along with two values for $\cs/(\cs_{{\rm std}}m_{{\rm DM}}[{\rm GeV}])$: 1 and 10. Here $\cs_{{\rm std}}=3\times10^{-26}$ cm$^{3}$s$^{-1}$ is the standard thermal cross section. For the meaning of each line, see the description given in the legend. The lowest dotted line represents the standard $\Lambda$CDM case with no additional energy input from the annihilating DM. {\bf (iii)} Matter temperature as a function of redshift. The models shown are exactly the same as in the panel above. The long dashed line compares the temperature of CMB. {\bf (iv)} The CMB visibility function $g(z)$ times redshift $z$ for the same models as shown in the above two panels. {\bf Righthand column} from top to bottom: {\bf (i)} Angular power spectra of CMB temperature fluctuations for the same models as already given above. {\bf (ii)} Temperature and E-mode polarization cross-spectra. {\bf (iii)} E-mode polarization spectra. In all of the righthand panels the points with errorbars show the 7-year measurements by the WMAP space mission.}
\label{fig2}
\end{figure*}

Because we are able to calculate $f$-parameters, we can go on to calculate the effect on cosmological recombination. To this end we modify the cosmological recombination code RECFAST \citep{1999ApJ...523L...1S} along the lines presented in \citet{2005PhRvD..72b3508P}. For more details see also \citet{2009A&A...505..999H}. Although more advanced cosmological recombination codes have recently been released\footnote{{\sc CosmoRec} (\url{www.Chluba.de/CosmoRec}) already provides a simple module that accounts for the effect of DM annihilation or more general energy injection, but this module is still being validated.} \citep{2010MNRAS.tmp.1876C, 2011PhRvD..83d3513A}, at this point we do not include any of the recently discuss corrections to the cosmological recombination process arising from detailed radiative transfer and atomic physics \citep[see][and references therein for overview]{2009ApJS..181..627F, 2010MNRAS.403..439R}.

The main effect of DM annihilations is a delay in recombination at $z\sim 1100$ and an increase in the low-redshift freeze-out tail. This changes the position and width of the last scattering surface and thus affects the CMB temperature and polarization anisotropies \citep[see, for example,][]{2004PhRvD..70d3502C, 2005PhRvD..72b3508P}. Since the input annihilation power scales as $\propto \rho_{DM}^2$, it is clear that structure formation leads to a significant boost over the average density squared $\bar{\rho}_{DM}^2$, i.e. $\langle \rho_{DM}^2(z)\rangle=B(z)\bar{\rho}_{DM}^2(z)$. The details of how we calculated the structure boost factors $B(z)$ can be found in \citet{2009A&A...505..999H}. The onset of structure formation at $z\sim 100$ is clearly visible in the upper lefthand panel of Fig.~\ref{fig2}. Here the three dark (light) solid curves, representing the results for the $\mu$ annihilation channel, correspond to $m_{{\rm DM}}=10$ GeV ($100$ GeV). From the bottom to top, the trio of lines in each case represent: (i) no structure formation, i.e. only a smooth background, (ii) including structures with the lower halo mass cutoff of $10^{-6}M_{\odot}$, (iii) structures with the cutoff of $10^{-9}M_{\odot}$. The halo mass function was assumed to have an analytic Sheth-Tormen form \citep{1999MNRAS.308..119S} and the halo mass-concentration relation followed \citet{2008MNRAS.391.1940M} description. For the full details of this calculations, again see \citet{2009A&A...505..999H}. The dotted lines represent the high-redshift fits (valid for $z\gtrsim 170$) for the $f$-parameters as given by \citet{2009PhRvD..80d3526S}. After adjusting slightly the high-redshift normalizations we see that the shape of their $f$-parameters at high redshifts agrees remarkably well with our results, even though our treatment is somewhat more simplified. The sharply peaked dashed curve at $z\sim 1000$ in Fig.~\ref{fig2} shows the Thomson visibility function \citep{1970Ap&SS...7....3S}. More specifically, the quantity $\mathcal{V}\sim z\, g(z){\rm d}\ln z$ gives the probability that the CMB photon last scattered in the logarithmic redshift interval ${\rm d}\ln z$ centered on redshift $z$. For the CMB calculations the values of the $f$-parameter matter only for the redshift range over which the visibility function is large, and in this range our results agree with the \citet{2009PhRvD..80d3526S} calculations to better than $5\%$ accuracy. We also tested the results for the $\tau$-channel and found very good agreement. Thus, our somewhat more simplistic calculation of the $f$-parameters compared to \citet{2009PhRvD..80d3526S} seems to be justified. 

The other lefthand panels of Fig.~\ref{fig2} (from top to bottom) show the fraction of free electrons, matter temperature, and the Thomson visibility function as a function of redshift for two DM particle masses, $10$ GeV and $100$ GeV, along with annihilation cross sections $\{10,100\}$ times, and $\{100,1000\}$ times the standard thermal cross section $\cs_{{\rm std}}=3\times10^{-26}$ cm$^{3}$s$^{-1}$. See the legend for the line definitions. The dotted lines show the standard behavior of ionization fraction, matter temperature and visibility function without any additional energy input from DM annihilation. In the panel for the matter temperature we also show the line for the temperature of the CMB. 

Because the annihilation power scales in proportion to $\cs n_{{\rm DM}}^2m_{{\rm DM}}\propto \cs/m_{{\rm DM}}$, we would expect the case with $m_{{\rm DM}}=10$ GeV, and $\cs=10\cs_{\rm std}$ ($100\cs_{\rm std}$) give comparable results to the case $m_{{\rm DM}}=100$ GeV and $\cs=100\cs_{\rm std}$ ($1000\cs_{\rm std}$), as long as the $f$-parameter does not depend strongly on $m_{{\rm DM}}$. This is indeed approximately so, as can be seen from Fig.~\ref{fig2}. Even though the lines for the ionization fraction, matter temperature, and CMB visibility function are easily separated at lower redshifts, this simple scaling with $\cs/m_{{\rm DM}}$ seems to hold very well near the peak of the visibility function, and so one would not expect to be able to clearly distinguish the models with the same value of $\cs/m_{{\rm DM}}$ via CMB measurements alone\footnote{To achieve better distinction between various models, one might try to use the differences in the behavior of the low-$z$ matter temperature and its observable consequence on the 21 cm transition measurements of the neutral hydrogen. The effect of annihilating DM on the 21 cm signal has been discussed e.g. in \citet{2006MNRAS.369.1719M,2006PhRvD..74j3502F,2007MNRAS.377..245V,2009PhRvD..80d3529N}.}. Also, the additional annihilation boost from the structure formation is not expected to influence the CMB measurements. Only in the very extreme case where instead of \citet{2008MNRAS.391.1940M} mass-concentration relation we use a simple power-law extrapolation down to the very low halo masses one is able to cause an additional significantly high low-redshift peak in the visibility function. However, for any realistic mass-concentration relation, along with annihilation cross sections $\cs$ that do not violate the CMB data, the contribution from the structure formation to the CMB signal is completely negligible. Also, the contribution to the low-redshift ionization fraction is very mild, and thus the annihilating DM models which are compatible with CMB measurements could only play a marginal role in reionizing the low-$z$ Universe. These results support our similar findings previously reported in \citet{2009A&A...505..999H}.

In the righthand panels of Fig.~\ref{fig2}, we show the temperature--temperature (TT), temperature--E-mode polarization (TE), and E-mode--E-mode (EE) angular power spectra. The points with error bars show the WMAP seven-year measurements \citep{2011ApJS..192...16L}. The dotted lines in all of the panels show theoretical predictions for the concordance $\Lambda$CDM model \citep{2011ApJS..192...18K}. As we might already expect, remembering the visibility function behavior as shown in the lowest lefthand panel, the models with the same values for $\cs/m_{{\rm DM}}$ are hardly distinguishable. This is because the redshift dependence of the $f$-parameter over the width of the visibility function is small, and so only the average value around $z\sim 1100$ really matters, in agreement with the statements of \citet{2009PhRvD..80d3526S}.

As a final note in this section we point out that, if one increases $\cs$ to high enough values and adds the structure formation boost, so that the low-redshift ionization fraction starts to get significantly approach one, the original RECFAST code, along with its DVERK solver, is not properly able to deal with the underlying set of stiff ODEs and simply breaks down. To be able to numerically treat these cases, one can use the significantly improved {\sc Recfast++} code\footnote{{\sc Recfast++} is part of more advanced recombination code {\sc CosmoRec} and can be downloaded from \url{www.Chluba.de/CosmoRec}.} \citep{2010MNRAS.tmp.1876C}, along with its stiff ODE solver. However, in our calculations the original RECFAST works fine, since the typically allowed values of $\cs/m_{{\rm DM}}$ compatible with the CMB data are low enough.

\section{CMB constraints}\label{sec4}

\begin{figure*}[ht!]
\centering
\includegraphics[width=\plotwdtwo]{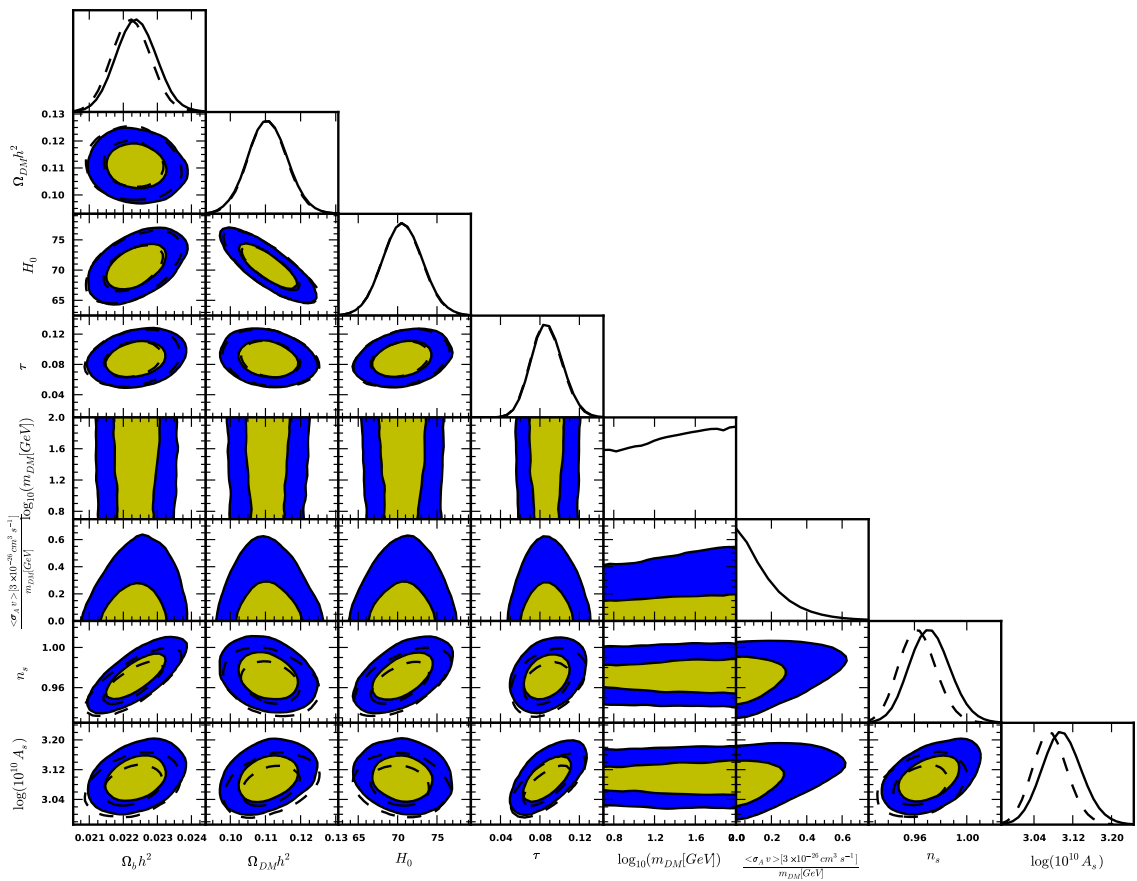}
\caption{The WMAP7 parameter constraints for the $\mu$ annihilation channel. Along with two annihilation parameters $\log_{10}(m_{{\rm DM}})$ and $\cs/m_{{\rm DM}}$, we also show the constraints on the other 6-parameter $\Lambda$CDM background model: $\Omega_{\rm B}h^2$, $\Omega_{\rm DM}h^2$, $H_0$, $\tau$, $n_{\rm S}$, $\log_{10}(A_{\rm S})$. From the inside out the colored 2D areas show the $1$-sigma and $2$-sigma regions after marginalization over the other parameters, and the dashed lines show the same regions for the basic 6-parameter $\Lambda$CDM without annihilating DM. The topmost panels in each column plot the marginalized 1D probability distributions for all of the parameters.}
\label{fig3}
\end{figure*}

Now we have all the ingredients available to calculate current and future CMB constraints on annihilating DM. We notice that the CMB constraints for the decaying DM are not competitive (e.g. when compared to the bounds available from the low-redshift gamma-ray measurements) due to $\propto \rho_{DM}$ scaling when compared to the $\propto \rho_{DM}^2$ scaling for the annihilating DM, and thus are not being discussed in this work.

For the currently existing CMB data, we used the latest power spectra, i.e. seven-year spectra, from the WMAP space mission \citep{2011ApJS..192...16L}. In reality the current CMB bounds can be somewhat tightened if one includes results from other smaller scale CMB measurements, in particular the measurements of the temperature power spectrum from ACBAR \citep{2009ApJ...694.1200R} and temperature and polarization spectra from QUaD \citep{2009ApJ...692.1247P}. In addition, data from ACT \citep{2011ApJ...729...62D} and SPT \citep{2010ApJ...719.1045L} could be added to the analysis, providing additional leverage on small scales\footnote{Since part of the effect of DM annihilation is degenerate with the effect of changing the spectral index of scalar perturbations, $n_{\rm S}$, small-scale CMB measurements help break this degeneracy.}. However, to keep our analysis clearer we decided to restrict ourselves to the WMAP7 data only. Regarding future CMB results we made predictions for the bounds available from the currently ongoing Planck mission, and also for the idealized noise-free experiment up to the multipole of $\ell_{\max}=2000$. As in \citet{2008arXiv0811.3918Z}, for Planck we assume $80\%$ sky coverage, beam size $\theta_{\rm FWHM}=7$ arcmin, and noisebias $N_{\ell}^{\rm TT}\simeq 10^{-4}$ $\mu$K$^2$ and $N_{\ell}^{\rm EE}\simeq 3.5\times10^{-4}$ $\mu$K$^2$ for TT and EE, respectively. For the ideal noise-free experiment, we assume a full sky coverage and the underlying uncertainties of the CMB fluctuations are solely due to a finite number of available fluctuation modes on the sky, i.e. due to cosmic variance, and so we call this type of idealized experiment `cosmic variance limited' (CVL) in the following. First steps towards CVL CMB measurements in both temperature and polarization down to small scales will become available from a combinations of {\sc SPTpol}\footnote{\url{http://pole.uchicago.edu/}} \citep{SPTpol} and {\sc ACTPol}\footnote{\url{http://www.physics.princeton.edu/act/}} \citep{ACTPol} along with Planck data, so that the CVL case discussed here provides a good guideline to what could become possible in the near future.

We generated the synthetic data for the Planck and CVL experiments as described in \url{http://cosmocoffee.info/}\footnote{See also \url{http://lpsc.in2p3.fr/perotto/} in case CMB lensing is needed.}. Along with WMAP7 likelihood code\footnote{\url{http://lambda.gsfc.nasa.gov/product/map/current/likelihood_info.cfm}}, they will be used in our Markov chain Monte Carlo (MCMC) parameter estimation. For the MCMC engine we used the publicly available CosmoMC tools\footnote{\url{http://cosmologist.info/cosmomc/}}\citep{2002PhRvD..66j3511L}, where the standard recombination modules were modified in order to allow additional energy input from annihilating DM. Our baseline concordance $\Lambda$CDM model \citep{2011ApJS..192...18K} is parameterized by six free parameters: $\Omega_{\rm B}h^2$, $\Omega_{\rm DM}h^2$, $H_0$, $\tau$, $n_{\rm S}$, $\log_{10}(A_{\rm S})$, where $\Omega_{\rm B}$ and $\Omega_{\rm DM}$ are density parameters for baryons and DM, $H_0=100\cdot h$ km/s/Mpc is the Hubble constant, $\tau$ is the reionization optical depth, and $A_{\rm S}$ and $n_{\rm S}$ are the amplitude and spectral index for adiabatic scalar perturbations, respectively. This minimal six-parameter set is extended by two additional parameters describing the annihilating DM: its mass $m_{{\rm DM}}$, and thermally averaged annihilation cross section $\cs$. Since MCMC will give better results if model parameters are as uncorrelated as possible, we used the parameter pair $\{m_{{\rm DM}},\cs/m_{{\rm DM}}\}$ in place of $\{m_{{\rm DM}},\cs\}$.

As an example, we show the WMAP7 constraints on all of the eight parameters, assuming $\mu$ annihilation channel on Fig.~\ref{fig3}. With dashed lines we show parameter constraints for the six-parameter $\Lambda$CDM model without annihilating DM. The colored 2D areas in all of the panels display 1-sigma and 2-sigma regions marginalized over all of the other parameters. For all of the parameters the topmost panels in each column plot the marginalized 1D probability distributions. It is reassuring to see that after introducing two additional parameters, all the previous six parameters, albeit with small shifts, are still as precisely determined. The strongest shifts compared to the baseline six-parameter model are quite understandably seen for parameters $A_{\rm S}$ and $n_{\rm S}$, which both lead to the rise of the CMB power spectra if increased, and thus counterbalance the damping of the spectrum caused by the extra scattering off the additional free electrons created by the energy input from the DM annihilation. 

Here it is important to note that these small shifts towards higher values of $A_{\rm S}$ and $n_{\rm S}$ are directed in the same direction as recently discussed recombination corrections from previously neglected standard physics processes \citep{2010MNRAS.403..439R, 2011arXiv1102.3683S}. Although biases in $A_{\rm S}$ and $n_{\rm S}$ arising from recombination physics are not significant for WMAP7 \citep{2011arXiv1102.3683S}, parameters derived from Planck data will be affected by several standard deviations if recombination corrections are neglected. This indicates that precise constraints on models with annihilating DM should probably account for these recombination corrections simultaneously.

\begin{figure}
\centering
\includegraphics[width=\plotwd]{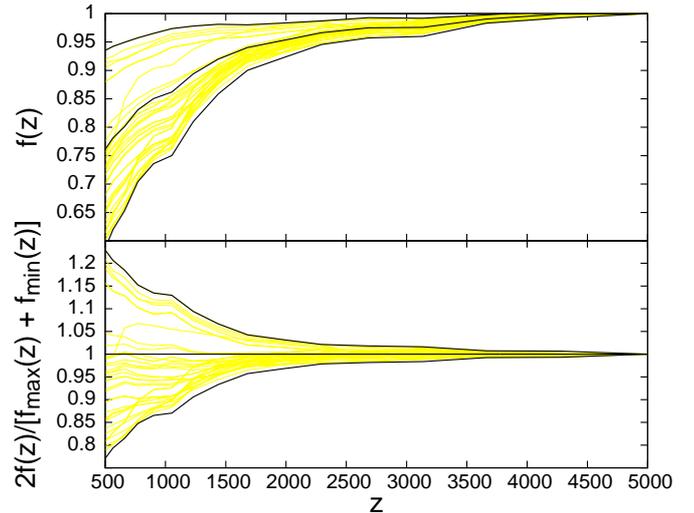}
\caption{{\bf Upper panel:} $f$-parameters for all of the channels considered in this work and for several DM particle masses in the interval $m_{{\rm DM}}=5-100$ GeV. For all of the cases, the asymptotic high-redshift $f$-parameter values have been renormalized to be equal to one. {\bf Lower panel:} The same as above, after dividing with a `typical shape', $(f_{\max}(z)+f_{\min}(z))/2$, of the $f$-parameter curve.}
\label{fig4}
\end{figure}

In Fig. \ref{fig3} we only present the constraints for one particular channel. However, we would like to provide easily usable recipes to calculate bounds for all of the channels described in Section~\ref{sec2}. As seen before, even though the $f$-parameters vary somewhat if one changes the mass from $10$ GeV to $100$ GeV, the CMB spectra as shown in Fig.~\ref{fig2} were fairly insensitive to this change as long as the ratio $\cs/m_{{\rm DM}}$ was kept the same. In the upper panel of Fig.~\ref{fig4} we plot the $f$-parameters for all of the channels and for several masses in the range $m_{{\rm DM}}=5-100$ GeV. Here we have renormalized the high-$z$ values to be equal to one. In the redshift range where the visibility function peaks (see the lowest lefthand panel of Fig.~\ref{fig2}), the variation among different DM models is not too large. To see this better, in the lower panel of Fig.~\ref{fig4} we have divided the above lines by a `typical shape' of the $f$-parameter given by $(f_{\max}(z)+f_{\min}(z))/2$, where $f_{\max}(z)$ and $f_{\min}(z)$ are the extremal values of $f(z)$ for a particular $z$. We see that, at most relevant redshifts, the variation among all the different models is around $15\%$.  This indeed suggests that it might be possible to provide a unified approximate scheme for calculating the bounds for all of the annihilation channels. 

\begin{figure}
\centering
\includegraphics[width=\plotwd]{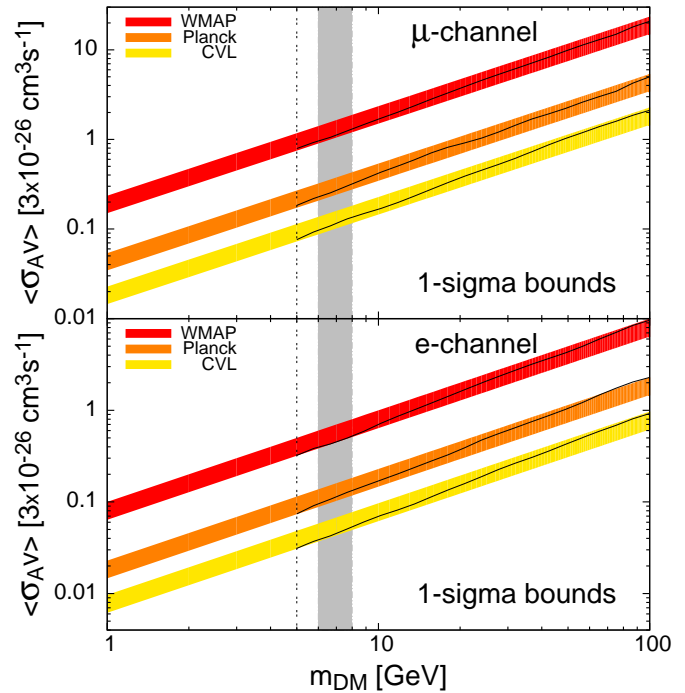}
\caption{WMAP, Planck, and CVL 1-sigma constraints on the $\cs-m_{{\rm DM}}$ plane for $\mu$ (upper panel) and $e$ (lower panel) annihilation channels. The solid lines show the upper bounds on annihilation cross section as determined directly through full MCMC calculations. The shaded regions around solid lines show the results from the simple recipe of Eq. (\ref{eq1}) with values of $r$ taken from Table~\ref{tab1}. The vertical gray stripe shows the range of WIMP masses ($m_{{\rm DM}}=6-8$ GeV) that provide a good fit to CoGeNT and DAMA/LIBRA data \citep{2010PhRvD..82l3509H}. The vertical dotted line marks the lowest DM particle mass $5$ GeV available for PYTHIA simulations. Extrapolations are shown below that value.}
\label{fig5}
\end{figure}

To bracket all the possibilities, we calculated the bounds assuming the $f$-parameter to have a typical shape $(f_{\max}(z)+f_{\min}(z))/2$ along with two extremal values $f_{\min}(z)$ and $f_{\max}(z)$. Of course, instead of two annihilation parameters we then only have to deal with one additional parameter: $\cs/m_{{\rm DM}}$. We took the asymptotic values for the $f$-parameter to have values of $0.25,0.50,0.75$, and $1.00$. This asymptotic value, as explained above, is simply $(1-f_{\nu})$, where $f_{\nu}$ is the fraction of the total energy carried away by neutrinos. It turned out that the resulting 1-sigma and 2-sigma constraints on $(1-f_{\nu})\cs/m_{{\rm DM}}$ for the above four values of $(1-f_{\nu})$ are the same with $(1-2)\%$ accuracy. Our final results (averaged over these small variations due to changing $(1-f_{\nu})$) can thus be given in the form

\begin{equation}\label{eq1}
(1-f_{\nu})\frac{\cs\,\left[3\times10^{-26}\,{\rm cm}^3{\rm s}^{-1}\right]}{m_{{\rm DM}}\,\left[{\rm GeV}\right]}< r\,,
\end{equation} 
where the values of $r$ for 1-sigma and 2-sigma bounds are given in Table~\ref{tab1}. One can see that the approximate bounds we provide are typically accurate at about the $20-30\%$ level. We tested that, indeed, the ranges of $r$ given in Table~\ref{tab1} fully cover the values of $r$ for all of the channels considered in this work. Two examples for the case of 1-sigma upper bounds are shown in Fig.~\ref{fig5}. Here the upper panel corresponds to the $\mu$-channel ($f_{\nu}\simeq 0.61$) and the lower one to the $e$-channel ($f_{\nu}\simeq 0.02$). We chose the above two channels because these are the two extreme cases among all of the channels treated in this paper. Consequently, they bracket the expected results for any realistic annihilating DM, given as a superposition over the basis channels. The solid lines show the bounds calculated directly through the MCMC analysis, while the shaded regions represent the ranges as obtained from Eq. (\ref{eq1}) and Table~\ref{tab1}. Indeed, the solid lines are fully covered by the shaded regions, as it should be. The vertical gray stripe indicates the range of WIMP masses ($m_{{\rm DM}}=6-8$ GeV) that provide a good fit to CoGeNT and DAMA/LIBRA data \citep{2010PhRvD..82l3509H}. The vertical dotted line marks the lowest DM particle mass $5$~GeV used in our PYTHIA simulations. This cut-off is not physical, but occurs because PYTHIA does not work below that energy. Therefore, the extrapolations of our results are shown below 5~GeV DM mass. Note that directly calculated lines for the $\mu$- and $e$-channel have slightly steeper slopes than given by the shaded regions, which increase as $\propto m_{{\rm DM}}$. As the $f$-parameters generally fall off more slowly for lower $m_{{\rm DM}}$, this behavior is also typical of the other channels. Thus, for lower $m_{{\rm DM}}$ values, one should actually get slightly stronger bounds than calculated directly from Eq. (\ref{eq1}), and so our extrapolations shown in Fig.~\ref{fig5} are somewhat conservative. 

\begin{table}
\centering
\caption{$1$-sigma and $2$-sigma values of $r$ in Eq. (\ref{eq1}) and their uncertainties for WMAP, Planck, and CVL CMB experiments.}
\label{tab1}
\begin{tabular}{l|l|l|}
 & \multicolumn{2}{c|}{$r$} \\
\cline{2-3}
 & {\footnotesize 1-sigma} & {\footnotesize 2-sigma} \\
\hline
WMAP & ${\bf {\scriptstyle 0.073}^{{\scriptscriptstyle +0.021}}_{{\scriptscriptstyle -0.013}}}$ & ${\bf {\scriptstyle 0.191}^{{\scriptscriptstyle +0.066}}_{{\scriptscriptstyle -0.031}}}$ \\
Planck & ${\bf {\scriptstyle 0.0160}^{{\scriptscriptstyle +0.0055}}_{{\scriptscriptstyle -0.0022}}}$ & ${\bf{\scriptstyle 0.0326}^{{\scriptscriptstyle +0.0096}}_{{\scriptscriptstyle -0.0055}}}$ \\
CVL & ${\bf {\scriptstyle 0.0071}^{{\scriptscriptstyle +0.0020}}_{{\scriptscriptstyle -0.0013}}}$ & ${\bf {\scriptstyle 0.0137}^{{\scriptscriptstyle +0.0031}}_{{\scriptscriptstyle -0.0032}}}$
\end{tabular}
\end{table}

We see that for WMAP7, depending on the annihilation model, the limiting DM particle mass below which the upper bound on the annihilation cross section drops below the standard thermal production value is in the range $4.5-10$ GeV, while the corresponding numbers reachable for the Planck and CVL experiment are $19-43$ GeV and $(45-100)$ GeV, respectively. Since the $\mu$-channel represents our most conservative case, one can instead say that, according to currently available CMB data, the annihilation cross section should be below the standard value of $3\times10^{-26}$ cm$^3$s$^{-1}$ as long as $m_{{\rm DM}}\lesssim 5$ GeV. \footnote{The numbers given here correspond to the 1-sigma upper bounds.} Thus, for the CoGeNT and DAMA/LIBRA best-fit mass region $6-8$ GeV, the standard thermal production cross section is still compatible with the CMB measurements. This could quite possibly be changed soon as Planck results become available. Of course one should keep in mind that the cross section bounds given here directly apply to redshifts of $z\sim 1000$, and if the cross section depends on velocity (as in the case of P-wave annihilation or Sommerfeld-enhanced scenario), one should be careful in converting these numbers to the values relevant at $z=0$. Also, note that the standard annihilation cross section $3\times10^{-26}$ cm$^3$s$^{-1}$, with respect to what we are comparing our CMB bounds to, provides the desired thermal relic density (i.e., $\Omega_{\rm DM}\sim 0.3$) only if WIMPs are annihilating through S-wave processes.

Except for a typical assumption of S-wave annihilation our results are largely model-independent, so all the particle-physics scenarios motivated by DAMA/LIBRA and CoGeNT results \citep{2010PhRvD..81k5005F,2010PhRvD..82d3522A,2010JCAP...08..018C,2010PhLB..692...65F,2010PhRvD..82c5019B,2010PhRvD..82l3509H,2010PhRvD..82g5004F,2010arXiv1004.0691E,2010PhRvD..82i5011B,2011PhRvD..83h3511C,2011PhLB..702..216B} including  theoretically well-motivated particle physics models that predict light DM, such as the MSSM \citep{2010PhRvD..81k7701F,2010PhRvD..81k1701K,2011PhRvD..83a5001F,2010PhRvD..81j7302B} and the NMSSM  \citep{2011JCAP...01..028K,2010arXiv1009.0549B,2010arXiv1009.2555G,2011PhRvL.106l1805D}, are stringently tested by WMAP and Planck.

A recent study by \citet{2011arXiv1102.5095D} suggests that annihilating DM with masses in the range $m_{{\rm DM}}=1-20$ GeV (i.e. compatible with CoGeNT and DAMA best-fit region), along with annihilation cross section $\cs\sim 9\times10^{-25}$ ${\rm cm}^3{\rm s}^{-1}$ (i.e. $30$ times the standard thermal relic cross section), may give a good match to the observed Fermi and WMAP haze. From Fig.~\ref{fig5} it is clear that, for the DM particle masses in this range, such high annihilation cross sections certainly conflict with available CMB data. If the required boost by a factor of $\sim 30$ is obtained via the Sommerfeld enhancement \citep{2009NuPhB.813....1C,2009PhRvD..79a5014A,2010JCAP...02..028S}, then at $z\sim 1000$ the expected cross section is at least as large as the value given above, so the model could be in even bigger trouble.  The alternative claims \citep{2009arXiv0910.2998G,2011PhLB..697..412H} of light DM annihilations in our Galaxy will also be stringently tested.

Finally, it would be interesting to compare the bounds given in \citet{2009PhRvD..80d3526S}, which  are based on the results obtained by \citet{2009PhRvD..80b3505G}, with our values. However, since \citet{2009PhRvD..80b3505G} used `on the spot' approximation for the energy deposition, the comparison can only be approximate. The WMAP5 2-sigma bound as given by Eq. (6) of \citet{2009PhRvD..80d3526S} can be cast in the form $\cs\,\left[3\times10^{-26}\,{\rm cm}^3{\rm s}^{-1}\right]/m_{{\rm DM}}\,\left[{\rm GeV}\right]<0.12/f$. To compare this with our WMAP7 results, we have to set $f_{\nu}=0$ in Eq. (\ref{eq1}) and, in the above relation, use a typical value for the $f$-parameter at $z\sim 1000$, which from Fig.~\ref{fig4} is $f\sim 0.85$. Thus, the value for $r$ that should be directly comparable to the 2-sigma WMAP value given in Table~\ref{tab1} is $0.12/0.85\sim 0.14$. Considering the differences in the treatment for the energy deposition, this value agrees reasonably well with our result. However, there are significantly greater differences if one compares the forecasts for the Planck and CVL experiments. In our case Planck and CVL would tighten the 2-sigma bound of WMAP by a factor of $\sim 6$ and $\sim 14$, respectively. The corresponding numbers ($\sim 13$ and $\sim 40$, respectively) from \citet{2009PhRvD..80b3505G} are certainly more optimistic. For CVL, some of this discrepancy is surely due to the higher $\ell_{\max}$ value assumed in \citet{2009PhRvD..80b3505G}: $\ell_{\max}=2500$ compared to our $\ell_{\max}=2000$. Even though we are comparing here the results derived from WMAP5 and WMAP7, the difference is expected to be negligible because the accuracy of the cosmological parameters with an additional two years of WMAP data improves only very moderately (typically somewhere between $5-15\%$), and thus cannot be the cause of the above discrepancy. 

NOTE: After the first version of this work was submitted, a new paper by \citet{2011PhRvD..84b7302G} appeared where the authors have performed a similar analysis (now also using WMAP7 data), this time finding results that agree with ours more closely: typical deviations are now within factors $1.2-2$, with their bounds on annihilation cross section being somewhat stronger. These relatively small deviations are actually quite negligible keeping in mind several approximations used in both analyses.

To see how much the omission of the additional excitations of atoms could change our results, we also performed several calculations where the exciting Ly-$\alpha$ photons are treated along the lines presented in \citet{2004PhRvD..70d3502C}. In agreement with the claim in \citet{2011PhRvD..84b7302G}, we find that the bounds on annihilation cross section are getting tighter only up to $\sim 10\%$; i.e., the excitations seem to have only a relatively weak effect with the current prescription.

\subsection*{Note on protons}
\label{sec:protons}
Although for most of the channels and, in particular, for the lower end of the considered DM particle mass range, the number of protons produced is completely negligible, for quark and gluon channels the contribution can reach $\sim 15\%$ of the total energy input if one has masses at the higher end of the considered range. As this is still only a relatively mild contribution, we did not attempt any detailed modeling for the proton component and simply assumed that all of this energy is absorbed as heat by the cosmic medium. A simple justification is the following. The energy loss rate for protons with the energies of interest in this paper is dominated by proton-proton scattering. Thus the loss rate $\Gamma\equiv -\frac{1}{E}\frac{{\rm d}E}{{\rm d}t}\simeq n_{\rm p}c\sigma_{\rm pp}K_{\rm pp}$, where $n_{\rm p}$ is the number density of target protons, $\sigma_{\rm pp}$ the scattering cross section, and $K_{\rm pp}$ the inelasticity parameter. The cross section $\sigma_{\rm pp}$ depends only weakly on proton energy with typical values of $30-40$ mbarn and inelasticity parameter $K_{\rm pp}\simeq 0.5$ for the energies of interest in this work (see e.g. \citealt{2008MNRAS.387..987W}). At redshifts $z\sim 1000$ we therefore get $\Gamma\simeq 10^{-13}$ s$^{-1}$ for the energy loss rate. Comparing this to the expansion rate $H$ at the same redshift, $H\simeq 0.4\times 10^{-13}$ s$^{-1}$, we see that $\Gamma \gtrsim H$, so one might expect a significant fraction of the proton energy to be absorbed. 

\section{Summary}

 In this paper we have calculated the existing and future CMB constraints on annihilating DM assuming the S-wave annihilation mode to be dominant. Our results can be summarized as follows.

\begin{itemize}
\item[$\bullet$] In full agreement with our earlier findings, first presented in \citet{2009A&A...505..999H}, we confirm that in case of any realistic halo mass-concentration relation, along with annihilation cross sections $\cs$ that do not violate the CMB data, the contribution from structure formation to the CMB signal is completely negligible. Also, the contribution to the low-redshift ionization fraction is very mild, and thus the annihilating DM models compatible with CMB measurements could only play a minor role in helping to reionize the low-$z$ Universe. Therefore for low-mass DM the CMB constraints can be considered
almost free of cosmological uncertainties. 
\item[$\bullet$]  At the same time, the particle physics uncertainties on the DM annihilation channels can all be described with one single parameter $f_\nu$, 
the energy fraction carried away by neutrinos. Therefore the DM annihilations to two muons represents the least stringently tested DM scenario.
\item[$\bullet$] As our main results, in the form of Eq. (\ref{eq1}) and Table~\ref{tab1}, we provided a simple recipe for estimating the upper bounds on annihilation cross section, valid for a broad range of DM models. Two examples for 1-sigma upper bounds for the $\mu$- and $e$-channels (which are the two extremal cases among all of the channels considered in this paper; i.e., those upper bounds bracket the expected results for more general annihilating DM models, given as a superposition over the basis channels) are shown in Fig.~\ref{fig5}.
\item[$\bullet$] For the DM particle masses $m_{{\rm DM}}=6-8$ GeV, which give best fits to CoGeNT and DAMA/LIBRA data, current CMB data is still compatible with a standard thermal relic annihilation cross section $3\times10^{-26}$ cm$^3$s$^{-1}$ only if the annihilations are dominantly into the $\mu$ or $\tau$ (and corresponding 4 lepton) channels. All other annihilation channels already now conflict with the CMB data.
\item[$\bullet$]  The sensitivity of  Planck space mission allows one to test all the DM annihilation channels definitively.  If the Planck mission will not find signals of annihilating DM in the CMB spectrum, all the light DM scenarios motivated by DAMA/LIBRA, CoGeNT, Fermi haze, and WMAP haze will be stringently tested if DM is a thermal relic.
\item[$\bullet$]  Our findings strongly disfavor a claim that annihilating DM with $m_{{\rm DM}}\sim 1-20$ GeV and $\cs\sim 9\times10^{-25}$ ${\rm cm}^3{\rm s}^{-1}$ (i.e. $\sim 30$ times above the standard value) could be a cause for Fermi and WMAP haze. 
\item[$\bullet$] The last conclusion applies to all DM scenarios with large boost factors from the Sommerfeld enhancement. Because CMB is sensitive to DM annihilations at $z\sim 1000$, at that time the Sommerfeld enhancement of annihilation cross section should have been  at least as large as today. Therefore, for light DM scenarios the boost of DM annihilation in our Galaxy can only come form the halo substructure.
\item[$\bullet$] Our MCMC calculations assumed standard 6-parameter $\Lambda$CDM cosmology, which was extended by two additional parameters describing the annihilating DM: $m_{{\rm DM}}$ and $\cs/m_{{\rm DM}}$. It turns out that, after introducing these 2 additional degrees of freedom, all the previous 6 parameters, albeit with small shifts (the most noticeable of which being the ones for $A_{\rm S}$ and $n_{\rm S}$), were still as precisely determined.
\item[$\bullet$] The weakest point of the analysis presented in this paper is the frequently used simple approximation about how the input energy of the cascading particles gets partitioned between ionizations and heating of the environment. For better treatment of this issue one possibly has to rely on detailed Monte Carlo calculations. This, however, we leave for a possible future study.  
\end{itemize}

\acknowledgements{We thank Marco Cirelli, Mario Kadastik, and Alessandro Strumia for discussions and our referee for useful comments and suggestions. GH acknowledges the support provided through a visiting scientist fellowship at the MPA. This work was supported by the ESF ERMOS Postdoctoral Research Grant 35, SF0060067s08, SF0690030s09, ESF8005, ESF8943, ESF8090, MTT8, MJD52, and by EU  FP7-INFRA-2007-1.2.3 contract No 223807. JC is also very grateful for additional financial support from the Beatrice~D.~Tremaine fellowship 2010.}
\bibliographystyle{aa}
\bibliography{references}

\begin{thebibliography}{97}
\expandafter\ifx\csname natexlab\endcsname\relax\def\natexlab#1{#1}\fi

\bibitem[{{Aalseth} {et~al.}(2011){Aalseth}, {Barbeau}, {Bowden},
  {Cabrera-Palmer}, {Colaresi}, {Collar}, {Dazeley}, {de Lurgio}, {Fast},
  {Fields}, {Greenberg}, {Hossbach}, {Keillor}, {Kephart}, {Marino}, {Miley},
  {Miller}, {Orrell}, {Radford}, {Reyna}, {Tench}, {van Wechel}, {Wilkerson},
  \& {Yocum}}]{2011PhRvL.106m1301A}
{Aalseth}, C.~E., {Barbeau}, P.~S., {Bowden}, N.~S., {et~al.} 2011, Physical
  Review Letters, 106, 131301

\bibitem[{{Abdo} {et~al.}(2009){Abdo}, {Ackermann}, {Ajello}, {Atwood},
  {Axelsson}, {Baldini}, {Ballet}, {Barbiellini}, {Bastieri}, {Battelino},
  {Baughman}, {Bechtol}, {Bellazzini}, {Berenji}, {Blandford}, {Bloom},
  {Bogaert}, {Bonamente}, {Borgland}, {Bregeon}, {Brez}, {Brigida}, {Bruel},
  {Burnett}, {Caliandro}, {Cameron}, {Caraveo}, {Carlson}, {Casandjian},
  {Cecchi}, {Charles}, {Chekhtman}, {Cheung}, {Chiang}, {Ciprini}, {Claus},
  {Cohen-Tanugi}, {Cominsky}, {Conrad}, {Cutini}, {Dermer}, {de Angelis}, {de
  Palma}, {Digel}, {di Bernardo}, {Do Couto E Silva}, {Drell}, {Dubois},
  {Dumora}, {Edmonds}, {Farnier}, {Favuzzi}, {Focke}, {Frailis}, {Fukazawa},
  {Funk}, {Fusco}, {Gaggero}, {Gargano}, {Gasparrini}, {Gehrels}, {Germani},
  {Giebels}, {Giglietto}, {Giordano}, {Glanzman}, {Godfrey}, {Grasso},
  {Grenier}, {Grondin}, {Grove}, {Guillemot}, {Guiriec}, {Hanabata}, {Harding},
  {Hartman}, {Hayashida}, {Hays}, {Hughes}, {J{\'o}hannesson}, {Johnson},
  {Johnson}, {Johnson}, {Kamae}, {Katagiri}, {Kataoka}, {Kawai}, {Kerr},
  {Kn{\"o}dlseder}, {Kocevski}, {Kuehn}, {Kuss}, {Lande}, {Latronico},
  {Lemoine-Goumard}, {Longo}, {Loparco}, {Lott}, {Lovellette}, {Lubrano},
  {Madejski}, {Makeev}, {Massai}, {Mazziotta}, {McConville}, {McEnery},
  {Meurer}, {Michelson}, {Mitthumsiri}, {Mizuno}, {Moiseev}, {Monte},
  {Monzani}, {Moretti}, {Morselli}, {Moskalenko}, {Murgia}, {Nolan}, {Norris},
  {Nuss}, {Ohsugi}, {Omodei}, {Orlando}, {Ormes}, {Ozaki}, {Paneque},
  {Panetta}, {Parent}, {Pelassa}, {Pepe}, {Pesce-Rollins}, {Piron}, {Pohl},
  {Porter}, {Profumo}, {Rain{\`o}}, {Rando}, {Razzano}, {Reimer}, {Reimer},
  {Reposeur}, {Ritz}, {Rochester}, {Rodriguez}, {Romani}, {Roth}, {Ryde},
  {Sadrozinski}, {Sanchez}, {Sander}, {Saz Parkinson}, {Scargle}, {Schalk},
  {Sellerholm}, {Sgr{\`o}}, {Smith}, {Smith}, {Spandre}, {Spinelli}, {Starck},
  {Stephens}, {Strickman}, {Strong}, {Suson}, {Tajima}, {Takahashi},
  {Takahashi}, {Tanaka}, {Thayer}, {Thayer}, {Thompson}, {Tibaldo}, {Tibolla},
  {Torres}, {Tosti}, {Tramacere}, {Uchiyama}, {Usher}, {van Etten},
  {Vasileiou}, {Vilchez}, {Vitale}, {Waite}, {Wallace}, {Wang}, {Winer},
  {Wood}, {Ylinen}, \& {Ziegler}}]{2009PhRvL.102r1101A}
{Abdo}, A.~A., {Ackermann}, M., {Ajello}, M., {et~al.} 2009, Physical Review
  Letters, 102, 181101

\bibitem[{{Adriani} {et~al.}(2009){Adriani}, {Barbarino}, {Bazilevskaya},
  {Bellotti}, {Boezio}, {Bogomolov}, {Bonechi}, {Bongi}, {Bonvicini}, {Bottai},
  {Bruno}, {Cafagna}, {Campana}, {Carlson}, {Casolino}, {Castellini}, {de
  Pascale}, {de Rosa}, {de Simone}, {di Felice}, {Galper}, {Grishantseva},
  {Hofverberg}, {Koldashov}, {Krutkov}, {Kvashnin}, {Leonov}, {Malvezzi},
  {Marcelli}, {Menn}, {Mikhailov}, {Mocchiutti}, {Orsi}, {Osteria}, {Papini},
  {Pearce}, {Picozza}, {Ricci}, {Ricciarini}, {Simon}, {Sparvoli},
  {Spillantini}, {Stozhkov}, {Vacchi}, {Vannuccini}, {Vasilyev}, {Voronov},
  {Yurkin}, {Zampa}, {Zampa}, \& {Zverev}}]{2009Natur.458..607A}
{Adriani}, O., {Barbarino}, G.~C., {Bazilevskaya}, G.~A., {et~al.} 2009, \nat,
  458, 607

\bibitem[{{Aharonian} {et~al.}(2009){Aharonian}, {Akhperjanian}, {Anton},
  {Barres de Almeida}, {Bazer-Bachi}, {Becherini}, {Behera}, {Bernl{\"o}hr},
  {Bochow}, {Boisson}, {Bolmont}, {Borrel}, {Brucker}, {Brun}, {Brun},
  {B{\"u}hler}, {Bulik}, {B{\"u}sching}, {Boutelier}, {Chadwick},
  {Charbonnier}, {Chaves}, {Cheesebrough}, {Chounet}, {Clapson}, {Coignet},
  {Dalton}, {Daniel}, {Davids}, {Degrange}, {Deil}, {Dickinson},
  {Djannati-Ata{\"i}}, {Domainko}, {O'C.~Drury}, {Dubois}, {Dubus}, {Dyks},
  {Dyrda}, {Egberts}, {Emmanoulopoulos}, {Espigat}, {Farnier}, {Feinstein},
  {Fiasson}, {F{\"o}rster}, {Fontaine}, {F{\"u}{\ss}ling}, {Gabici}, {Gallant},
  {G{\'e}rard}, {Gerbig}, {Giebels}, {Glicenstein}, {Gl{\"u}ck}, {Goret},
  {G{\"o}ring}, {Hauser}, {Hauser}, {Heinz}, {Heinzelmann}, {Henri}, {Hermann},
  {Hinton}, {Hoffmann}, {Hofmann}, {Holleran}, {Hoppe}, {Horns},
  {Jacholkowska}, {de Jager}, {Jahn}, {Jung}, {Katarzy{\'n}ski}, {Katz},
  {Kaufmann}, {Kendziorra}, {Kerschhaggl}, {Khangulyan}, {Kh{\'e}lifi},
  {Keogh}, {Klu{\'z}niak}, {Kneiske}, {Komin}, {Kosack}, {Kossakowski},
  {Lamanna}, {Lenain}, {Lohse}, {Marandon}, {Martin}, {Martineau-Huynh},
  {Marcowith}, {Masbou}, {Maurin}, {McComb}, {Medina}, {Moderski}, {Moulin},
  {Naumann-Godo}, {de Naurois}, {Nedbal}, {Nekrassov}, {Nicholas}, {Niemiec},
  {Nolan}, {Ohm}, {Olive}, {de O{\~n}a Wilhelmi}, {Orford}, {Ostrowski},
  {Panter}, {Paz Arribas}, {Pedaletti}, {Pelletier}, {Petrucci}, {Pita},
  {P{\"u}hlhofer}, {Punch}, {Quirrenbach}, {Raubenheimer}, {Raue}, {Rayner},
  {Reimer}, {Renaud}, {Rieger}, {Ripken}, {Rob}, {Rosier-Lees}, {Rowell},
  {Rudak}, {Rulten}, {Ruppel}, {Sahakian}, {Santangelo}, {Schlickeiser},
  {Sch{\"o}ck}, {Schr{\"o}der}, {Schwanke}, {Schwarzburg}, {Schwemmer},
  {Shalchi}, {Sikora}, {Skilton}, {Sol}, {Spangler}, {Stawarz}, {Steenkamp},
  {Stegmann}, {Stinzing}, {Superina}, {Szostek}, {Tam}, {Tavernet}, {Terrier},
  {Tibolla}, {Tluczykont}, {van Eldik}, {Vasileiadis}, {Venter}, {Venter},
  {Vialle}, {Vincent}, {Vivier}, {V{\"o}lk}, {Volpe}, {Wagner}, {Ward},
  {Zdziarski}, \& {Zech}}]{2009A&A...508..561A}
{Aharonian}, F., {Akhperjanian}, A.~G., {Anton}, G., {et~al.} 2009, \aap, 508,
  561

\bibitem[{{Ahmed} {et~al.}(2011){Ahmed}, {Akerib}, {Arrenberg}, {Bailey},
  {Balakishiyeva}, {Baudis}, {Bauer}, {Brink}, {Bruch}, {Bunker}, {Cabrera},
  {Caldwell}, {Cooley}, {Do Couto E Silva}, {Cushman}, {Daal}, {Dejongh}, {di
  Stefano}, {Dragowsky}, {Duong}, {Fallows}, {Figueroa-Feliciano}, {Filippini},
  {Fox}, {Fritts}, {Golwala}, {Hall}, {Hennings-Yeomans}, {Hertel}, {Holmgren},
  {Hsu}, {Huber}, {Kamaev}, {Kiveni}, {Kos}, {Leman}, {Liu}, {Mahapatra},
  {Mandic}, {McCarthy}, {Mirabolfathi}, {Moore}, {Nelson}, {Ogburn}, {Phipps},
  {Pyle}, {Qiu}, {Ramberg}, {Rau}, {Reisetter}, {Resch}, {Saab}, {Sadoulet},
  {Sander}, {Schnee}, {Seitz}, {Serfass}, {Sundqvist}, {Tarka}, {Wikus},
  {Yellin}, {Yoo}, {Young}, \& {Zhang}}]{2011PhRvL.106m1302A}
{Ahmed}, Z., {Akerib}, D.~S., {Arrenberg}, S., {et~al.} 2011, Physical Review
  Letters, 106, 131302

\bibitem[{{Ali-Ha{\"i}moud} \& {Hirata}(2011)}]{2011PhRvD..83d3513A}
{Ali-Ha{\"i}moud}, Y. \& {Hirata}, C.~M. 2011, \prd, 83, 043513

\bibitem[{{Andreas} {et~al.}(2010){Andreas}, {Arina}, {Hambye}, {Ling}, \&
  {Tytgat}}]{2010PhRvD..82d3522A}
{Andreas}, S., {Arina}, C., {Hambye}, T., {Ling}, F., \& {Tytgat}, M.~H.~G.
  2010, \prd, 82, 043522

\bibitem[{{Angle} {et~al.}(2009){Angle}, {Aprile}, {Arneodo}, {Baudis},
  {Bernstein}, {Bolozdynya}, {Coelho}, {Dahl}, {Deviveiros}, {Ferella},
  {Fernandes}, {Fiorucci}, {Gaitskell}, {Giboni}, {Gomez}, {Hasty}, {Kastens},
  {Kwong}, {Lopes}, {Madden}, {Manalaysay}, {Manzur}, {McKinsey}, {Monzani},
  {Ni}, {Oberlack}, {Orboeck}, {Plante}, {Santorelli}, {Dos Santos}, {Shagin},
  {Shutt}, {Sorensen}, {Schulte}, {Winant}, \&
  {Yamashita}}]{2009PhRvD..80k5005A}
{Angle}, J., {Aprile}, E., {Arneodo}, F., {et~al.} 2009, \prd, 80, 115005

\bibitem[{{Aprile} {et~al.}(2010){Aprile}, {Arisaka}, {Arneodo}, {Askin},
  {Baudis}, {Behrens}, {Bokeloh}, {Brown}, {Cardoso}, {Choi}, {Cline},
  {Fattori}, {Ferella}, {Giboni}, {Kish}, {Lam}, {Lamblin}, {Lang}, {Lim},
  {Lopes}, {Marrod{\'a}n Undagoitia}, {Mei}, {Melgarejo Fernandez}, {Ni},
  {Oberlack}, {Orrigo}, {Pantic}, {Plante}, {Ribeiro}, {Santorelli}, {Dos
  Santos}, {Schumann}, {Shagin}, {Teymourian}, {Thers}, {Tziaferi}, {Wang},
  {Weinheimer}, \& {XENON100 Collaboration}}]{2010PhRvL.105m1302A}
{Aprile}, E., {Arisaka}, K., {Arneodo}, F., {et~al.} 2010, Physical Review
  Letters, 105, 131302

\bibitem[{{Arina} \& {Tytgat}(2011)}]{2011JCAP...01..011A}
{Arina}, C. \& {Tytgat}, M.~H.~G. 2011, \jcap, 1, 11

\bibitem[{{Arkani-Hamed} {et~al.}(2009){Arkani-Hamed}, {Finkbeiner}, {Slatyer},
  \& {Weiner}}]{2009PhRvD..79a5014A}
{Arkani-Hamed}, N., {Finkbeiner}, D.~P., {Slatyer}, T.~R., \& {Weiner}, N.
  2009, \prd, 79, 015014

\bibitem[{{Barger} {et~al.}(2010{\natexlab{a}}){Barger}, {Gao}, {McCaskey}, \&
  {Shaughnessy}}]{2010PhRvD..82i5011B}
{Barger}, V., {Gao}, Y., {McCaskey}, M., \& {Shaughnessy}, G.
  2010{\natexlab{a}}, \prd, 82, 095011

\bibitem[{{Barger} {et~al.}(2009){Barger}, {Keung}, {Marfatia}, \&
  {Shaughnessy}}]{2009PhLB..672..141B}
{Barger}, V., {Keung}, W., {Marfatia}, D., \& {Shaughnessy}, G. 2009, Physics
  Letters B, 672, 141

\bibitem[{{Barger} {et~al.}(2010{\natexlab{b}}){Barger}, {McCaskey}, \&
  {Shaughnessy}}]{2010PhRvD..82c5019B}
{Barger}, V., {McCaskey}, M., \& {Shaughnessy}, G. 2010{\natexlab{b}}, \prd,
  82, 035019

\bibitem[{{Baxter} {et~al.}(2010){Baxter}, {Dodelson}, {Koushiappas}, \&
  {Strigari}}]{2010PhRvD..82l3511B}
{Baxter}, E.~J., {Dodelson}, S., {Koushiappas}, S.~M., \& {Strigari}, L.~E.
  2010, \prd, 82, 123511

\bibitem[{{Belikov} {et~al.}(2010){Belikov}, {Gunion}, {Hooper}, \&
  {Tait}}]{2010arXiv1009.0549B}
{Belikov}, A.~V., {Gunion}, J.~F., {Hooper}, D., \& {Tait}, T.~M.~P. 2010,
  arXiv:1009.0549

\bibitem[{{Bell} \& {Jacques}(2009)}]{2009PhRvD..79d3507B}
{Bell}, N.~F. \& {Jacques}, T.~D. 2009, \prd, 79, 043507

\bibitem[{{Bergstr{\"o}m} {et~al.}(2009){Bergstr{\"o}m}, {Bertone},
  {Bringmann}, {Edsj{\"o}}, \& {Taoso}}]{2009PhRvD..79h1303B}
{Bergstr{\"o}m}, L., {Bertone}, G., {Bringmann}, T., {Edsj{\"o}}, J., \&
  {Taoso}, M. 2009, \prd, 79, 081303

\bibitem[{{Bergstr{\"o}m} {et~al.}(2008){Bergstr{\"o}m}, {Bringmann}, \&
  {Edsj{\"o}}}]{2008PhRvD..78j3520B}
{Bergstr{\"o}m}, L., {Bringmann}, T., \& {Edsj{\"o}}, J. 2008, \prd, 78, 103520

\bibitem[{{Bernabei} {et~al.}(2010){Bernabei}, {Belli}, {Cappella}, {Cerulli},
  {Dai}, {D'Angelo}, {He}, {Incicchitti}, {Kuang}, {Ma}, {Montecchia},
  {Nozzoli}, {Prosperi}, {Sheng}, {Wang}, \& {Ye}}]{2010EPJC...67...39B}
{Bernabei}, R., {Belli}, P., {Cappella}, F., {et~al.} 2010, European Physical
  Journal C, 67, 39

\bibitem[{{Bertone} {et~al.}(2009){Bertone}, {Cirelli}, {Strumia}, \&
  {Taoso}}]{2009JCAP...03..009B}
{Bertone}, G., {Cirelli}, M., {Strumia}, A., \& {Taoso}, M. 2009, \jcap, 3, 9

\bibitem[{{Bertone} {et~al.}(2005){Bertone}, {Hooper}, \&
  {Silk}}]{2005PhR...405..279B}
{Bertone}, G., {Hooper}, D., \& {Silk}, J. 2005, \physrep, 405, 279

\bibitem[{{Bottino} {et~al.}(2010){Bottino}, {Donato}, {Fornengo}, \&
  {Scopel}}]{2010PhRvD..81j7302B}
{Bottino}, A., {Donato}, F., {Fornengo}, N., \& {Scopel}, S. 2010, \prd, 81,
  107302

\bibitem[{{Buckley} {et~al.}(2011){Buckley}, {Hooper}, \&
  {Tait}}]{2011PhLB..702..216B}
{Buckley}, M.~R., {Hooper}, D., \& {Tait}, T.~M.~P. 2011, Physics Letters B,
  702, 216

\bibitem[{{Chang} {et~al.}(2010){Chang}, {Liu}, {Pierce}, {Weiner}, \&
  {Yavin}}]{2010JCAP...08..018C}
{Chang}, S., {Liu}, J., {Pierce}, A., {Weiner}, N., \& {Yavin}, I. 2010, \jcap,
  8, 18

\bibitem[{{Chen} \& {Kamionkowski}(2004)}]{2004PhRvD..70d3502C}
{Chen}, X. \& {Kamionkowski}, M. 2004, \prd, 70, 043502

\bibitem[{{Chluba}(2010)}]{2010MNRAS.402.1195C}
{Chluba}, J. 2010, \mnras, 402, 1195

\bibitem[{{Chluba} \& {Thomas}(2010)}]{2010MNRAS.tmp.1876C}
{Chluba}, J. \& {Thomas}, R.~M. 2010, \mnras, 1876

\bibitem[{{Cirelli} {et~al.}(2011){Cirelli}, {Corcella}, {Hektor}, {H{\"u}tsi},
  {Kadastik}, {Panci}, {Raidal}, {Sala}, \& {Strumia}}]{2011JCAP...03..051C}
{Cirelli}, M., {Corcella}, G., {Hektor}, A., {et~al.} 2011, \jcap, 3, 51

\bibitem[{{Cirelli} {et~al.}(2009{\natexlab{a}}){Cirelli}, {Iocco}, \&
  {Panci}}]{2009JCAP...10..009C}
{Cirelli}, M., {Iocco}, F., \& {Panci}, P. 2009{\natexlab{a}}, \jcap, 10, 9

\bibitem[{{Cirelli} {et~al.}(2009{\natexlab{b}}){Cirelli}, {Kadastik},
  {Raidal}, \& {Strumia}}]{2009NuPhB.813....1C}
{Cirelli}, M., {Kadastik}, M., {Raidal}, M., \& {Strumia}, A.
  2009{\natexlab{b}}, Nuclear Physics B, 813, 1

\bibitem[{{Cirelli} \& {Panci}(2009)}]{2009NuPhB.821..399C}
{Cirelli}, M. \& {Panci}, P. 2009, Nuclear Physics B, 821, 399

\bibitem[{{Cirelli} {et~al.}(2010){Cirelli}, {Panci}, \&
  {Serpico}}]{2010NuPhB.840..284C}
{Cirelli}, M., {Panci}, P., \& {Serpico}, P.~D. 2010, Nuclear Physics B, 840,
  284

\bibitem[{{Cline} {et~al.}(2011){Cline}, {Frey}, \&
  {Chen}}]{2011PhRvD..83h3511C}
{Cline}, J.~M., {Frey}, A.~R., \& {Chen}, F. 2011, \prd, 83, 083511

\bibitem[{{Das} {et~al.}(2011){Das}, {Marriage}, {Ade}, {Aguirre}, {Amiri},
  {Appel}, {Barrientos}, {Battistelli}, {Bond}, {Brown}, {Burger}, {Chervenak},
  {Devlin}, {Dicker}, {Bertrand Doriese}, {Dunkley}, {D{\"u}nner},
  {Essinger-Hileman}, {Fisher}, {Fowler}, {Hajian}, {Halpern}, {Hasselfield},
  {Hern{\'a}ndez-Monteagudo}, {Hilton}, {Hilton}, {Hincks}, {Hlozek},
  {Huffenberger}, {Hughes}, {Hughes}, {Infante}, {Irwin}, {Baptiste Juin},
  {Kaul}, {Klein}, {Kosowsky}, {Lau}, {Limon}, {Lin}, {Lupton}, {Marsden},
  {Martocci}, {Mauskopf}, {Menanteau}, {Moodley}, {Moseley}, {Netterfield},
  {Niemack}, {Nolta}, {Page}, {Parker}, {Partridge}, {Reid}, {Sehgal},
  {Sherwin}, {Sievers}, {Spergel}, {Staggs}, {Swetz}, {Switzer}, {Thornton},
  {Trac}, {Tucker}, {Warne}, {Wollack}, \& {Zhao}}]{2011ApJ...729...62D}
{Das}, S., {Marriage}, T.~A., {Ade}, P.~A.~R., {et~al.} 2011, \apj, 729, 62

\bibitem[{{Dobler} {et~al.}(2011){Dobler}, {Cholis}, \&
  {Weiner}}]{2011arXiv1102.5095D}
{Dobler}, G., {Cholis}, I., \& {Weiner}, N. 2011, arXiv:1102.5095

\bibitem[{{Dobler} \& {Finkbeiner}(2008)}]{2008ApJ...680.1222D}
{Dobler}, G. \& {Finkbeiner}, D.~P. 2008, \apj, 680, 1222

\bibitem[{{Dobler} {et~al.}(2010){Dobler}, {Finkbeiner}, {Cholis}, {Slatyer},
  \& {Weiner}}]{2010ApJ...717..825D}
{Dobler}, G., {Finkbeiner}, D.~P., {Cholis}, I., {Slatyer}, T., \& {Weiner}, N.
  2010, \apj, 717, 825

\bibitem[{{Draper} {et~al.}(2011){Draper}, {Liu}, {Wagner}, {Wang}, \&
  {Zhang}}]{2011PhRvL.106l1805D}
{Draper}, P., {Liu}, T., {Wagner}, C.~E.~M., {Wang}, L.-T., \& {Zhang}, H.
  2011, Physical Review Letters, 106, 121805

\bibitem[{{Einasto}(2009)}]{2009arXiv0901.0632E}
{Einasto}, J. 2009, arXiv:0901.0632

\bibitem[{{Essig} {et~al.}(2010){Essig}, {Kaplan}, {Schuster}, \&
  {Toro}}]{2010arXiv1004.0691E}
{Essig}, R., {Kaplan}, J., {Schuster}, P., \& {Toro}, N. 2010, arXiv:1004.0691

\bibitem[{{Feldman} {et~al.}(2010){Feldman}, {Liu}, \&
  {Nath}}]{2010PhRvD..81k7701F}
{Feldman}, D., {Liu}, Z., \& {Nath}, P. 2010, \prd, 81, 117701

\bibitem[{{Fendt} {et~al.}(2009){Fendt}, {Chluba}, {Rubi{\~n}o-Mart{\'{\i}}n},
  \& {Wandelt}}]{2009ApJS..181..627F}
{Fendt}, W.~A., {Chluba}, J., {Rubi{\~n}o-Mart{\'{\i}}n}, J.~A., \& {Wandelt},
  B.~D. 2009, \apjs, 181, 627

\bibitem[{{Feng}(2010)}]{2010ARA&A..48..495F}
{Feng}, J.~L. 2010, \araa, 48, 495

\bibitem[{{Fitzpatrick} {et~al.}(2010){Fitzpatrick}, {Hooper}, \&
  {Zurek}}]{2010PhRvD..81k5005F}
{Fitzpatrick}, A.~L., {Hooper}, D., \& {Zurek}, K.~M. 2010, \prd, 81, 115005

\bibitem[{{Fitzpatrick} \& {Zurek}(2010)}]{2010PhRvD..82g5004F}
{Fitzpatrick}, A.~L. \& {Zurek}, K.~M. 2010, \prd, 82, 075004

\bibitem[{{Foot}(2010)}]{2010PhLB..692...65F}
{Foot}, R. 2010, Physics Letters B, 692, 65

\bibitem[{{Fornengo} {et~al.}(2011){Fornengo}, {Scopel}, \&
  {Bottino}}]{2011PhRvD..83a5001F}
{Fornengo}, N., {Scopel}, S., \& {Bottino}, A. 2011, \prd, 83, 015001

\bibitem[{{Fox} \& {Poppitz}(2009)}]{2009PhRvD..79h3528F}
{Fox}, P.~J. \& {Poppitz}, E. 2009, \prd, 79, 083528

\bibitem[{{Furlanetto} {et~al.}(2006){Furlanetto}, {Oh}, \&
  {Pierpaoli}}]{2006PhRvD..74j3502F}
{Furlanetto}, S.~R., {Oh}, S.~P., \& {Pierpaoli}, E. 2006, \prd, 74, 103502

\bibitem[{{Galli} {et~al.}(2009){Galli}, {Iocco}, {Bertone}, \&
  {Melchiorri}}]{2009PhRvD..80b3505G}
{Galli}, S., {Iocco}, F., {Bertone}, G., \& {Melchiorri}, A. 2009, \prd, 80,
  023505

\bibitem[{{Galli} {et~al.}(2011){Galli}, {Iocco}, {Bertone}, \&
  {Melchiorri}}]{2011PhRvD..84b7302G}
{Galli}, S., {Iocco}, F., {Bertone}, G., \& {Melchiorri}, A. 2011, \prd, 84,
  027302

\bibitem[{{Goodenough} \& {Hooper}(2009)}]{2009arXiv0910.2998G}
{Goodenough}, L. \& {Hooper}, D. 2009, arXiv:0910.2998

\bibitem[{{Gunion} {et~al.}(2010){Gunion}, {Belikov}, \&
  {Hooper}}]{2010arXiv1009.2555G}
{Gunion}, J.~F., {Belikov}, A.~V., \& {Hooper}, D. 2010, arXiv:1009.2555

\bibitem[{{Hooper} {et~al.}(2010){Hooper}, {Collar}, {Hall}, {McKinsey}, \&
  {Kelso}}]{2010PhRvD..82l3509H}
{Hooper}, D., {Collar}, J.~I., {Hall}, J., {McKinsey}, D., \& {Kelso}, C.~M.
  2010, \prd, 82, 123509

\bibitem[{{Hooper} \& {Goodenough}(2011)}]{2011PhLB..697..412H}
{Hooper}, D. \& {Goodenough}, L. 2011, Physics Letters B, 697, 412

\bibitem[{{H{\"u}tsi} {et~al.}(2009){H{\"u}tsi}, {Hektor}, \&
  {Raidal}}]{2009A&A...505..999H}
{H{\"u}tsi}, G., {Hektor}, A., \& {Raidal}, M. 2009, \aap, 505, 999

\bibitem[{{H{\"u}tsi} {et~al.}(2010){H{\"u}tsi}, {Hektor}, \&
  {Raidal}}]{2010JCAP...07..008H}
{H{\"u}tsi}, G., {Hektor}, A., \& {Raidal}, M. 2010, \jcap, 7, 8

\bibitem[{{Jungman} {et~al.}(1996){Jungman}, {Kamionkowski}, \&
  {Griest}}]{1996PhR...267..195J}
{Jungman}, G., {Kamionkowski}, M., \& {Griest}, K. 1996, \physrep, 267, 195

\bibitem[{{Kang} {et~al.}(2011){Kang}, {Li}, {Liu}, {Tong}, \&
  {Yang}}]{2011JCAP...01..028K}
{Kang}, Z., {Li}, T., {Liu}, T., {Tong}, C., \& {Yang}, J.~M. 2011, \jcap, 1,
  28

\bibitem[{{Kanzaki} {et~al.}(2010){Kanzaki}, {Kawasaki}, \&
  {Nakayama}}]{2010PThPh.123..853K}
{Kanzaki}, T., {Kawasaki}, M., \& {Nakayama}, K. 2010, Progress of Theoretical
  Physics, 123, 853

\bibitem[{{Komatsu} {et~al.}(2011){Komatsu}, {Smith}, {Dunkley}, {Bennett},
  {Gold}, {Hinshaw}, {Jarosik}, {Larson}, {Nolta}, {Page}, {Spergel},
  {Halpern}, {Hill}, {Kogut}, {Limon}, {Meyer}, {Odegard}, {Tucker}, {Weiland},
  {Wollack}, \& {Wright}}]{2011ApJS..192...18K}
{Komatsu}, E., {Smith}, K.~M., {Dunkley}, J., {et~al.} 2011, \apjs, 192, 18

\bibitem[{{Kopp} {et~al.}(2010){Kopp}, {Schwetz}, \&
  {Zupan}}]{2010JCAP...02..014K}
{Kopp}, J., {Schwetz}, T., \& {Zupan}, J. 2010, \jcap, 2, 14

\bibitem[{{Kuflik} {et~al.}(2010){Kuflik}, {Pierce}, \&
  {Zurek}}]{2010PhRvD..81k1701K}
{Kuflik}, E., {Pierce}, A., \& {Zurek}, K.~M. 2010, \prd, 81, 111701

\bibitem[{{Larson} {et~al.}(2011){Larson}, {Dunkley}, {Hinshaw}, {Komatsu},
  {Nolta}, {Bennett}, {Gold}, {Halpern}, {Hill}, {Jarosik}, {Kogut}, {Limon},
  {Meyer}, {Odegard}, {Page}, {Smith}, {Spergel}, {Tucker}, {Weiland},
  {Wollack}, \& {Wright}}]{2011ApJS..192...16L}
{Larson}, D., {Dunkley}, J., {Hinshaw}, G., {et~al.} 2011, \apjs, 192, 16

\bibitem[{{Lewis} \& {Bridle}(2002)}]{2002PhRvD..66j3511L}
{Lewis}, A. \& {Bridle}, S. 2002, \prd, 66, 103511

\bibitem[{{Lueker} {et~al.}(2010){Lueker}, {Reichardt}, {Schaffer}, {Zahn},
  {Ade}, {Aird}, {Benson}, {Bleem}, {Carlstrom}, {Chang}, {Cho}, {Crawford},
  {Crites}, {de Haan}, {Dobbs}, {George}, {Hall}, {Halverson}, {Holder},
  {Holzapfel}, {Hrubes}, {Joy}, {Keisler}, {Knox}, {Lee}, {Leitch}, {McMahon},
  {Mehl}, {Meyer}, {Mohr}, {Montroy}, {Padin}, {Plagge}, {Pryke}, {Ruhl},
  {Shaw}, {Shirokoff}, {Spieler}, {Stalder}, {Staniszewski}, {Stark},
  {Vanderlinde}, {Vieira}, \& {Williamson}}]{2010ApJ...719.1045L}
{Lueker}, M., {Reichardt}, C.~L., {Schaffer}, K.~K., {et~al.} 2010, \apj, 719,
  1045

\bibitem[{{Macci{\`o}} {et~al.}(2008){Macci{\`o}}, {Dutton}, \& {van den
  Bosch}}]{2008MNRAS.391.1940M}
{Macci{\`o}}, A.~V., {Dutton}, A.~A., \& {van den Bosch}, F.~C. 2008, \mnras,
  391, 1940

\bibitem[{{Mapelli} {et~al.}(2006){Mapelli}, {Ferrara}, \&
  {Pierpaoli}}]{2006MNRAS.369.1719M}
{Mapelli}, M., {Ferrara}, A., \& {Pierpaoli}, E. 2006, \mnras, 369, 1719

\bibitem[{{McMahon} {et~al.}(2009){McMahon}, {Aird}, {Benson}, {Bleem},
  {Britton}, {Carlstrom}, {Chang}, {Cho}, {de Haan}, {Crawford}, \&
  {Crites}}]{SPTpol}
{McMahon}, J.~J., {Aird}, K.~A., {Benson}, B.~A., {et~al.} 2009, in American
  Institute of Physics Conference Series, Vol. 1185, American Institute of
  Physics Conference Series, ed. {B.~Young, B.~Cabrera, \& A.~Miller}, 511--514

\bibitem[{{Meade} {et~al.}(2010){Meade}, {Papucci}, {Strumia}, \&
  {Volansky}}]{2010NuPhB.831..178M}
{Meade}, P., {Papucci}, M., {Strumia}, A., \& {Volansky}, T. 2010, Nuclear
  Physics B, 831, 178

\bibitem[{{Natarajan} \& {Schwarz}(2009)}]{2009PhRvD..80d3529N}
{Natarajan}, A. \& {Schwarz}, D.~J. 2009, \prd, 80, 043529

\bibitem[{{Niemack} {et~al.}(2010){Niemack}, {Ade}, {Aguirre}, {Barrientos},
  {Beall}, {Bond}, {Britton}, {Cho}, {Das}, {Devlin}, {Dicker}, \&
  {Dunkley}}]{ACTPol}
{Niemack}, M.~D., {Ade}, P.~A.~R., {Aguirre}, J., {et~al.} 2010, in Presented
  at the Society of Photo-Optical Instrumentation Engineers (SPIE) Conference,
  Vol. 7741, Society of Photo-Optical Instrumentation Engineers (SPIE)
  Conference Series

\bibitem[{{Padmanabhan} \& {Finkbeiner}(2005)}]{2005PhRvD..72b3508P}
{Padmanabhan}, N. \& {Finkbeiner}, D.~P. 2005, \prd, 72, 023508

\bibitem[{{Papucci} \& {Strumia}(2010)}]{2010JCAP...03..014P}
{Papucci}, M. \& {Strumia}, A. 2010, \jcap, 3, 14

\bibitem[{{Peebles} {et~al.}(2000){Peebles}, {Seager}, \&
  {Hu}}]{2000ApJ...539L...1P}
{Peebles}, P.~J.~E., {Seager}, S., \& {Hu}, W. 2000, \apjl, 539, L1

\bibitem[{{Pryke} {et~al.}(2009){Pryke}, {Ade}, {Bock}, {Bowden}, {Brown},
  {Cahill}, {Castro}, {Church}, {Culverhouse}, {Friedman}, {Ganga}, {Gear},
  {Gupta}, {Hinderks}, {Kovac}, {Lange}, {Leitch}, {Melhuish}, {Memari},
  {Murphy}, {Orlando}, {Schwarz}, {O'Sullivan}, {Piccirillo}, {Rajguru},
  {Rusholme}, {Taylor}, {Thompson}, {Turner}, {Wu}, \&
  {Zemcov}}]{2009ApJ...692.1247P}
{Pryke}, C., {Ade}, P., {Bock}, J., {et~al.} 2009, \apj, 692, 1247

\bibitem[{{Reichardt} {et~al.}(2009){Reichardt}, {Ade}, {Bock}, {Bond},
  {Brevik}, {Contaldi}, {Daub}, {Dempsey}, {Goldstein}, {Holzapfel}, {Kuo},
  {Lange}, {Lueker}, {Newcomb}, {Peterson}, {Ruhl}, {Runyan}, \&
  {Staniszewski}}]{2009ApJ...694.1200R}
{Reichardt}, C.~L., {Ade}, P.~A.~R., {Bock}, J.~J., {et~al.} 2009, \apj, 694,
  1200

\bibitem[{{Rubi{\~n}o-Mart{\'{\i}}n} {et~al.}(2010){Rubi{\~n}o-Mart{\'{\i}}n},
  {Chluba}, {Fendt}, \& {Wandelt}}]{2010MNRAS.403..439R}
{Rubi{\~n}o-Mart{\'{\i}}n}, J.~A., {Chluba}, J., {Fendt}, W.~A., \& {Wandelt},
  B.~D. 2010, \mnras, 403, 439

\bibitem[{{Savage} {et~al.}(2011){Savage}, {Gelmini}, {Gondolo}, \&
  {Freese}}]{2011PhRvD..83e5002S}
{Savage}, C., {Gelmini}, G., {Gondolo}, P., \& {Freese}, K. 2011, \prd, 83,
  055002

\bibitem[{{Schwetz}(2010)}]{2010arXiv1011.5432S}
{Schwetz}, T. 2010, arXiv:1011.5432

\bibitem[{{Seager} {et~al.}(1999){Seager}, {Sasselov}, \&
  {Scott}}]{1999ApJ...523L...1S}
{Seager}, S., {Sasselov}, D.~D., \& {Scott}, D. 1999, \apjl, 523, L1

\bibitem[{{Shaw} \& {Chluba}(2011)}]{2011arXiv1102.3683S}
{Shaw}, J.~R. \& {Chluba}, J. 2011, \mnras, 415, 1343

\bibitem[{{Sheth} \& {Tormen}(1999)}]{1999MNRAS.308..119S}
{Sheth}, R.~K. \& {Tormen}, G. 1999, \mnras, 308, 119

\bibitem[{{Shull} \& {van Steenberg}(1985)}]{1985ApJ...298..268S}
{Shull}, J.~M. \& {van Steenberg}, M.~E. 1985, \apj, 298, 268

\bibitem[{{Sj{\"o}strand} {et~al.}(2008){Sj{\"o}strand}, {Mrenna}, \&
  {Skands}}]{2008CoPhC.178..852S}
{Sj{\"o}strand}, T., {Mrenna}, S., \& {Skands}, P. 2008, Computer Physics
  Communications, 178, 852

\bibitem[{{Slatyer}(2010)}]{2010JCAP...02..028S}
{Slatyer}, T.~R. 2010, \jcap, 2, 28

\bibitem[{{Slatyer} {et~al.}(2009){Slatyer}, {Padmanabhan}, \&
  {Finkbeiner}}]{2009PhRvD..80d3526S}
{Slatyer}, T.~R., {Padmanabhan}, N., \& {Finkbeiner}, D.~P. 2009, \prd, 80,
  043526

\bibitem[{{Sunyaev} \& {Zeldovich}(1970)}]{1970Ap&SS...7....3S}
{Sunyaev}, R.~A. \& {Zeldovich}, Y.~B. 1970, \apss, 7, 3

\bibitem[{{Vald{\'e}s} {et~al.}(2007){Vald{\'e}s}, {Ferrara}, {Mapelli}, \&
  {Ripamonti}}]{2007MNRAS.377..245V}
{Vald{\'e}s}, M., {Ferrara}, A., {Mapelli}, M., \& {Ripamonti}, E. 2007,
  \mnras, 377, 245

\bibitem[{{Vincent} {et~al.}(2010){Vincent}, {Xue}, \&
  {Cline}}]{2010PhRvD..82l3519V}
{Vincent}, A.~C., {Xue}, W., \& {Cline}, J.~M. 2010, \prd, 82, 123519

\bibitem[{{Wommer} {et~al.}(2008){Wommer}, {Melia}, \&
  {Fatuzzo}}]{2008MNRAS.387..987W}
{Wommer}, E., {Melia}, F., \& {Fatuzzo}, M. 2008, \mnras, 387, 987

\bibitem[{{XENON100 Collaboration} {et~al.}(2011){XENON100 Collaboration},
  {Aprile}, {Arisaka}, {Arneodo}, {Askin}, {Baudis}, {Behrens}, {Bokeloh},
  {Brown}, {Bruch}, {Cardoso}, {Choi}, {Cline}, {Duchovni}, {Fattori},
  {Ferella}, {Giboni}, {Gross}, {Kish}, {Lam}, {Lamblin}, {Lang}, {Lim},
  {Lindemann}, {Lindner}, {Lopes}, {Marrod{\'a}n Undagoitia}, {Mei}, {Melgarejo
  Fernandez}, {Ni}, {Oberlack}, {Orrigo}, {Pantic}, {Plante}, {Ribeiro},
  {Santorelli}, {dos Santos}, {Schumann}, {Shagin}, {Teymourian}, {Thers},
  {Tziaferi}, {Vitells}, {Wang}, {Weber}, \&
  {Weinheimer}}]{2011arXiv1103.0303X}
{XENON100 Collaboration}, {Aprile}, E., {Arisaka}, K., {et~al.} 2011,
  arXiv:1103.0303

\bibitem[{{Zaldarriaga} {et~al.}(2008){Zaldarriaga}, {Colombo}, {Komatsu},
  {Lidz}, {Mortonson}, {Oh}, {Pierpaoli}, {Verde}, \&
  {Zahn}}]{2008arXiv0811.3918Z}
{Zaldarriaga}, M., {Colombo}, L., {Komatsu}, E., {et~al.} 2008, arXiv:0811.3918

\bibitem[{{Zavala} {et~al.}(2011){Zavala}, {Vogelsberger}, {Slatyer}, {Loeb},
  \& {Springel}}]{2011PhRvD..83l3513Z}
{Zavala}, J., {Vogelsberger}, M., {Slatyer}, T.~R., {Loeb}, A., \& {Springel},
  V. 2011, \prd, 83, 123513

\bibitem[{{Zdziarski} \& {Svensson}(1989)}]{1989ApJ...344..551Z}
{Zdziarski}, A.~A. \& {Svensson}, R. 1989, \apj, 344, 551

\bibitem[{{Zhang} {et~al.}(2006){Zhang}, {Chen}, {Lei}, \&
  {Si}}]{2006PhRvD..74j3519Z}
{Zhang}, L., {Chen}, X., {Lei}, Y.-A., \& {Si}, Z.-G. 2006, \prd, 74, 103519

\end{thebibliography}

\end{document}